\documentclass[pmlr]{jmlr}

\RequirePackage{graphicx}
 \usepackage{booktabs,siunitx,array}
\usepackage{graphicx}   
\usepackage{csquotes}
\usepackage{float}
\usepackage{subcaption}
\usepackage{wrapfig}
\usepackage{longtable}
\makeatletter
\def\set@curr@file#1{\def\@curr@file{#1}} 
\makeatother
\usepackage[load-configurations=version-1]{siunitx} 

\theorembodyfont{\upshape}
\theoremheaderfont{\scshape}
\theorempostheader{:}
\theoremsep{\newline}

\jmlrvolume{[VOLUME \# 340]}
\jmlryear{2026}
\jmlrworkshop{Machine Learning for Healthcare}

\title[Causal Discovery Radiation]{Causal Discovery of Radiation Response Mechanisms in Human Cells}

\author{\Name{Ashka Shah}
       \Email{shahashka@uchicago.edu}\\ 
       \addr Department of Computer Science\\
       University of Chicago\\
       Chicago, IL, USA 
       \AND
       \Name{Rick Stevens}
       \Email{stevens@cs.uchicago.edu}\\ 
       \addr Department of Computer Science\\
       University of Chicago\\
       Chicago, IL, USA} 

\begin{document}

\maketitle

\begin{abstract}
Next-generation sequencing technologies, including RNA-sequencing, provide genome-wide measurements of gene expression and enable broad explorations of biomarkers and mechanisms underlying disease and treatment response. Bioinformatics tools for processing this data, such as differential expression analysis, are largely univariate, linear, and rely on predefined pathway knowledge annotations, which limits their ability to capture nonlinear and multivariate gene interactions. This paper explores the application of causal discovery to characterizing transcriptional responses to radiation as a function of dose rate in human cells. By jointly modeling radiation perturbations and gene expression, we learn directed gene networks that capture important regulatory relationships beyond correlation and exhibit significant enrichment of known radiation response pathways compared to baseline approaches. We find that inferred causal graphs reveal structured network features such as high in-degree housekeeping genes and high out-degree transcription factors. Further analysis suggests a hierarchical organization of stress response pathways and triggered cell death pathways. This work highlights the potential of causal discovery in healthcare settings with applications to understanding response mechanisms, identifying regulatory targets, and improving interpretation of complex genomic data. Code is available at \texttt{https://github.com/shahashka/lucid\_cd}.
\end{abstract}

\section{Introduction}
\label{sec:introduction} 

 RNA-sequencing experiments provide genome-wide insight into gene expression levels across conditions at high resolution; however, finding genes of interest from this vast data remains a difficult challenge \citep{trapnell2009tophat,love2014moderated}. Differential expression analysis is the most popular tool used to identify genes that statistically vary between perturbed and control samples as an initial step to identify important genes \citep{rosati2024differential}. Subsequent pathway enrichment onto knowledge bases is used to understand the molecular mechanisms underlying the conditions analyzed. Finally, biomarkers are found by linking genes to known disease pathways. Despite its popularity, differential expression analysis combined with pathway enrichment suffers from two main drawbacks: (1) genes are identified using linear models and univariate tests; this misses coordinated and nonlinear signals from genes that may individually be weak (2) its outputs do not directly imply any mechanistic understanding limiting discovery of novel mechanisms and pathways. To address these issues multivariate tests \citep{glazko2009unite}, supervised machine learning with feature ranking \citep{wenric2018using}, and graph learning algorithms \citep{marbach2012wisdom, moerman2019grnboost2} have been applied to identify important genes under perturbed conditions. 
 
 Recently causal discovery (algorithms for inferring cause and effect relationships represented by graphs) has emerged as a data-driven approach to mechanism discovery. 
 Causal discovery often focuses on synthetic data because real world data violate the assumptions needed for identifiability of the true causal graph; these include the absence of confounding variables, acyclic relationships, and large sample sizes. There is a growing call for the application of these methods to more realistic and scientific settings \citep{brouillard2024landscape}. To this end, benchmarking frameworks with large-scale transcriptomics and CRISPR perturbation datasets have been developed; however, they largely reveal that causal discovery algorithms fail to learn regulatory relationships between genes from single-cell RNA-sequencing datasets \citep{chevalley2025large}. 

\begin{figure}
\centering
    \includegraphics[width=0.7\linewidth]{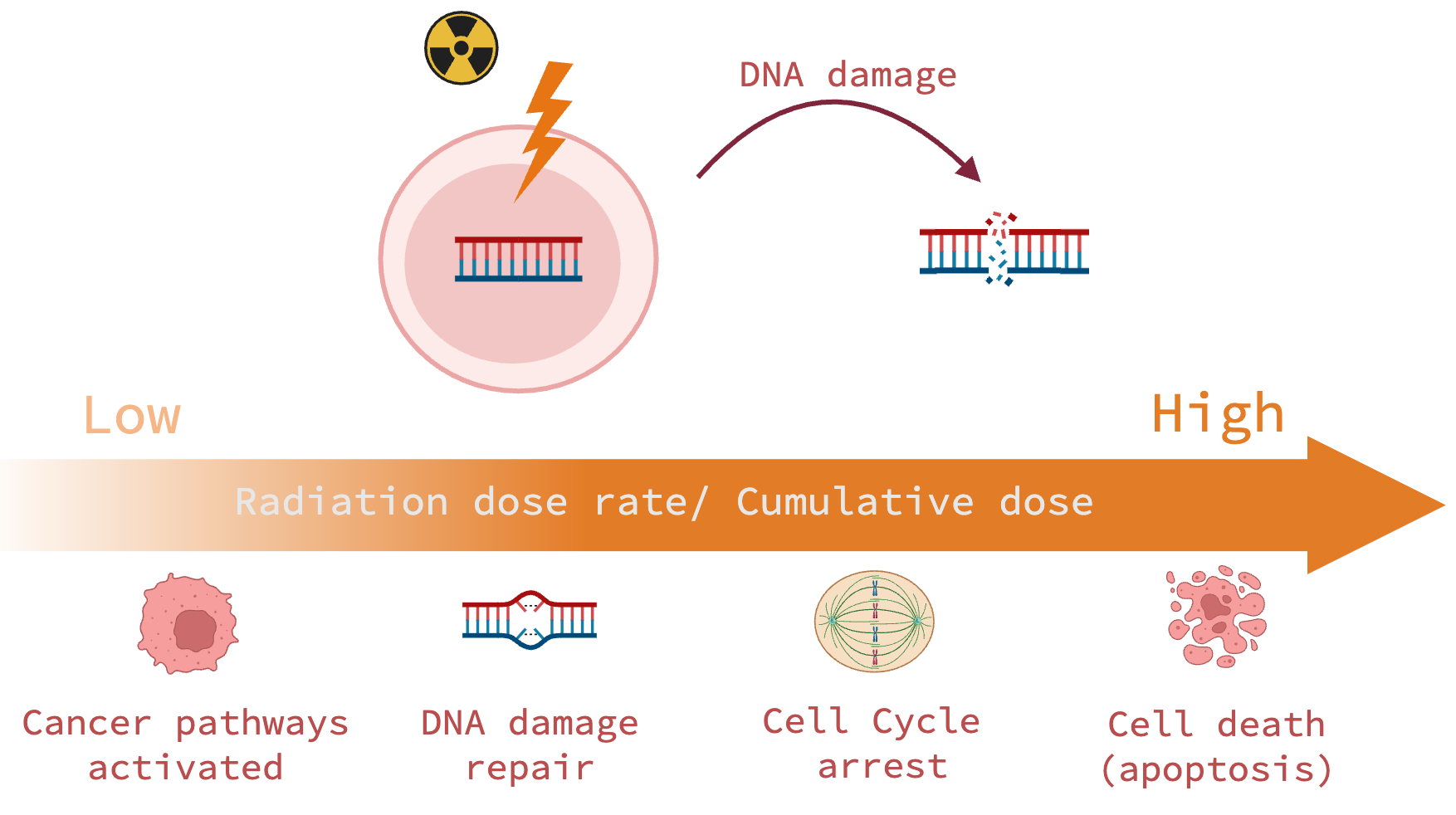}
   \caption{Exposure to ionizing radiation causes DNA breakage in cells. Cells activate pathways in order to repair the damage before the cell divides. When damage is excessive or occurs too rapidly, cells initiate programmed cell death (apoptosis).} 
   \label{fig:radiation-response}
   \end{figure}

In this paper, we take an alternative approach by applying causal discovery to low-sample, perturbed-control transcriptomics studies for understanding radiation response. We jointly model radiation exposure, which is viewed as a perturbation upstream of gene regulation that affects many pathways, alongside the multivariate nonlinear distribution of gene expression. We apply this framework to RNA sequencing data sampled from RPE1 cell lines exposed to chronic low-dose radiation across increasing dose rates. Such prolonged exposure is associated with elevated risks of cancer and cardiovascular disease and is known to activate cellular stress-response pathways \citep{shimizu2010radiation}. A complete understanding of the interplay between dose rate, cumulative dose, and cell type in the activation of low-dose radiation pathways remains an open question (Fig. \ref{fig:radiation-response}) and has implications for people chronically exposed to low-dose radiation (e.g., space exploration, radon exposure) \citep{verheij2000radiation}. To address this gap, we use causal discovery to uncover relationships linking radiation exposure to gene expression, and provide mechanistic insight into how dose rate affects cellular stress-response pathways.

\subsection*{Generalizable Insights about Machine Learning in the Context of Healthcare}
This work applies causal discovery to infer important genes and to learn novel pathways from high-dimensional biological data with limited samples and structured perturbations, a common setting across healthcare (e.g., drug response, disease progression, and other treatment-driven processes). The limitations of traditional pipelines, such as differential expression analysis, are ubiquitous in these settings: overly simplistic univariate and linear model assumptions, and a lack of discovery power by relying on incomplete knowledge bases. We demonstrate that causal discovery can be a complementary method that considers nonlinearity, multivariate interactions, and the distribution of the perturbation variable to identify driver variables and provide structure-based hypotheses of novel mechanisms.  

\section{Related Work}
\subsection{Finding Important Genes: Univariate and Multivariate Approaches }
 Differential expression analysis is the standard bioinformatics tool used to infer genes with expression levels that differ significantly between conditions (e.g., perturbed and control). This analysis tests the null hypothesis that the logarithmic fold change (LFC) between perturbed and control for a gene’s expression is exactly zero, i.e., that the gene is not affected by the perturbation. The goal is to produce a list of genes passing multiple-test adjustment, ranked by $p$-value. 

The most popular tool \texttt{DESeq2} \citep{love2014moderated} employs this approach; each gene is modeled by a generalized linear model and negative binomial distribution. \texttt{DESEq2} models relationships between genes by assuming that genes of similar average expression have a similar dispersion; however, the statistical test for determining significant change between condition groups is univariate for each gene. Therefore, this method does not account for correlations and coordinated effects between groups of genes. 

Multivariate tests including Hotelling's $T^2$ \citep{lu2005hotelling} and $N$-statistic \citep{klebanov2007multivariate} consider the joint distribution of expression levels. \citet{glazko2009unite} evaluate univariate and multivariate tests and show that different tests find common but also complementing differentially expressed gene sets -- each test projects on different aspects of the data. They propose use of both univariate and multivariate tests for the analysis of biological data for increased power.

\citet{wenric2018using} propose multivariate classifiers to jointly model effects across genes and classify samples into perturbed and control groups. They find that random forest classifiers outperform differential expression analysis in 9 of the 12 cancer datasets evaluated. 
They further demonstrate that differential expression analysis might miss important genes, and a supervised learning-based gene selection method can be used to supplement these shortcomings.

\subsection{Inferring Novel Pathways with Graph Learning}
Once important genes are identified pathway enrichment is used to identify the processes that underlie response to perturbations \citep{raudvere2019g}. Pathway enrichment algorithms determine the significance of gene overlap to known pathways, recorded in knowledge bases like Gene Ontology \citep{gene2019gene}, KEGG \citep{kanehisa2002kegg}, Reactome \citep{milacic2024reactome}, and WikiPathways \citep{agrawal2024wikipathways}. 

There is growing interest in developing algorithms that can identify novel pathways to guide the discovery of new mechanisms and biomarkers. Many gene regulatory network (GRN) inference algorithms have been proposed to solve this challenge. These methods include GENIE3 \citep{marbach2012wisdom} which uses an ensemble of random forest regression models to recover transcription factor (TF) to target gene directed relationships. GRNBoost2 \citep{moerman2019grnboost2} takes a similar approach, but with gradient boosting which is much faster and more scalable. 
 Methods like GENELink \citep{chen2022graph} and GRINDC \citep{feng2023gene} use deep learning to predict the presence or absence of an edge from embedded representations of single-cell RNA-sequencing datasets \citep{dixit2016perturb}. Several of these algorithms have been benchmarked against BEELINE \citep{pratapa2020benchmarking}. The authors show that GRN inference remains an open problem despite these recent advances.  

Causal discovery methods infer directed acyclic graphs (DAGs) from data. These approaches benefit from theoretical guarantees of identifying the true causal graph under certain assumptions. 
A breakthrough in causal discovery was the definition of a continuous acyclicity constraint, allowing for gradient-based learning of DAGs \citep{zheng2018dags}. This includes deep learning approaches such as DAG-GNN \citep{yu2019dag} which employs a variational autoencoder to learn an adjacency matrix under the DAG constraint while capturing nonlinear relationships between genes. 
A recent benchmark paper \citep{chevalley2025large} evaluates several causal discovery algorithms with single-cell RNA-seq and Perturb-seq datasets, using knowledge graphs for edge validation. Their findings underscored issues with scalability and suboptimal use of perturbational data, and report low precision and recall of ground truth regulatory edges.


In this paper, we use causal discovery to learn causal graphs over the joint distribution of gene expression \textit{and a perturbation variable}. Therefore, we learn pathways specific to perturbation response. Our goal is not to reconstruct all known regulatory relationships, which previous works have shown is notoriously difficult, but rather to propose an alternative approach for identifying important genes under perturbed conditions. A key advantage of our approach is the inference of a directed graph over variables, enabling the generation of hypotheses about novel response mechanisms. To our knowledge this is the first use of causal discovery in this setting. 

\section{Data Collection}
\label{sec:data}
The bulk RNA-seq dataset is described by \cite{jantre2026interpretable}. Here, we provide a brief overview of the procedure. RPE1 cells were grown in the lab and treated to low-dose ionizing radiation from a Cesium-137 gamma ray source at $D=5$ different dose rates measured in milliGray (mGy) per hour \footnote{One Gray is equal to one joule of absorbed radiation energy per kilogram of matter}, including a control group with no radiation exposure. RNA-seq measurements were made for control and treated samples after each week for $T=9$ weeks with 2 replicates per group for a total of $n=2\times (D+1)\times T=108$ samples. Raw RNA-seq reads were preprocessed with alignment to the human reference genome and filtered for low quality reads. Additionally, transcripts per million (TPM) values were computed per sample using \texttt{TPMCalculator v0.0.3}, resulting in a data matrix $X_{\text{TPM}} \in \mathbb{R}^{n \times p}$, where $p=15,694$ genes. We denote $x_{(t,d),g}$ as a value in the matrix corresponding to week $t$, dose rate $d$  (including $d=0$ controls) and gene $g$. We compute the log2-Fold-Change matrix $X_{\text{LFC}} \in \mathbb{R}^{m \times p}$, where $m=T \times D=45$ treated conditions. This matrix is computed element-wise for each gene : $X_{{\text{LFC}}_{(t,d),g}} = \text{log}_2\Big(\frac{x_{(t,d),g}}{x_{(t,0)g}}\Big)$.

\section{Methods}
\begin{figure}
\centering
    \includegraphics[width=0.8\linewidth]{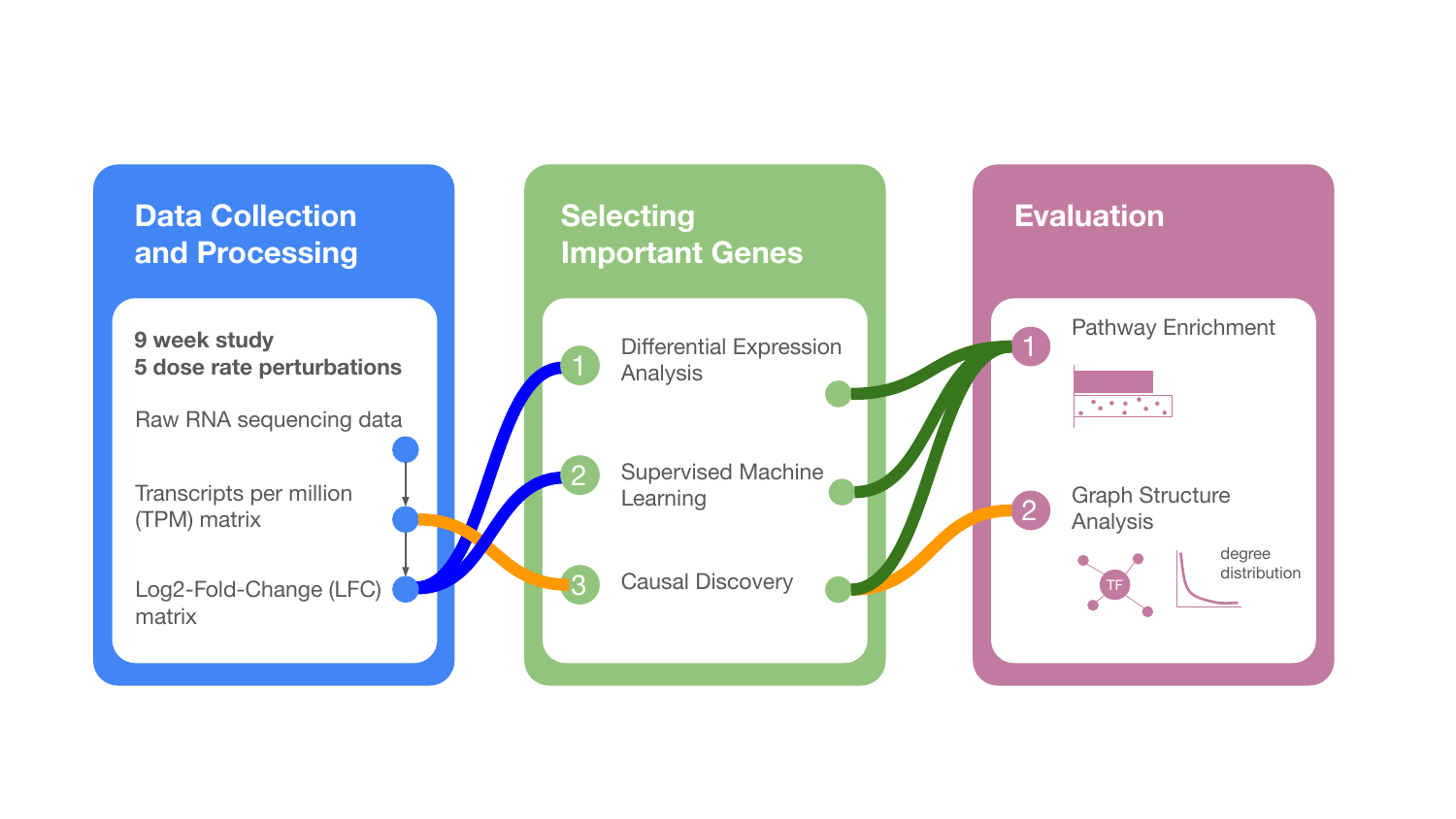}
    \caption{Our workflow comprises three main pipelines. Our contribution, shown in orange, is the use of causal discovery to identify important genes from transcriptomics data and downstream graph structural evaluation. We compare against standard pipelines for selecting important genes and subsequent pathway enrichment.}
    \label{fig:workflow}
\end{figure}

A visualization of our framework is shown in Fig. \ref{fig:workflow} which is comprised of three pipelines. In this section, we describe the second and third pipelines.
\label{sec:methods}
\subsection{Selecting Important Genes}
\subsubsection{Differential Expression Analysis}
Differential expression analysis was performed using \texttt{pyDESeq2 v0.4.9} with dataset $X_{\text{LFC}}$ to determine which genes had statistically significant expression change in each of the 45 control-exposed pairs. Wald tests were performed and $p$-values were adjusted for multiple testing using the Benjamini–Hochberg false discovery rate (FDR). Genes with adjusted $p\text{-value} < 0.05$ and $| X_{\text{LFC}_{(t,d),g}}| > 1$ were considered differentially expressed. For analysis, we aggregate differentially expressed genes (DEGs) at each dose rate $d \in D$ across time: $G_d = \underset{t \in T}{\cup}\{ g: |X_{\text{LFC}_{(t,d),g}}| > 1, p\text{-value} <0.05\}$. 
\subsubsection{Supervised Machine Learning}
\texttt{RandomForestRegression} models were trained to predict the \texttt{(week,dose\_rate)} labels of each sample in $X_{\text{LFC}}$. Weeks were assigned ordinal values $T={1,2,3,4,5,6,7,8,9}$ and dose rates were assigned $D=0.28, 0.38, 0.55, 6.66, 12.11$ (mGy/hr) scalar values according to the measured exposure. We perform \texttt{RepeatedKFold} cross validation with 5 folds and 5 repeated runs. Within each fold, data is normalized to between (0,1]. The top 1000 features in each trained model's \texttt{feature\_importance\_} vector were extracted, and we selected the final set of genes that were stable in 20 out of 25 folds. Note that this gene set is dose agnostic since we use a stratified training split of \texttt{(week, dose\_rate)} labels to train the models. 

\subsubsection{Causal Discovery}
We use \texttt{DAG-GNN} \citep{yu2019dag}, a variational autoencoder parameterized by a graph neural network, to construct DAGs at each dose rate from $X_\text{TPM}$ and an additional column vector with the cumulative radiation of each sample, computed as \texttt{dose\_rate} $\times$ \texttt{week} $\times$ 168 hours/week. We split $X_\text{TPM}$ row-wise by dose rate so that $X_{\text{TPM}_{d}}$ represents the samples collected at dose rate $d$. \texttt{DAG-GNN} training is time and memory intensive; to learn an adjacency matrix over
$O(10,000)$ variables would be intractable. Therefore, we first filter out genes at each dose rate with $p\text{-value} > 0.05$ from the differential expression analysis, thereby removing genes that were not significantly expressed while still retaining genes with small LFC values. The resultant $X_{\text{TPM}_{d}}$'s at each dose rate after this preprocessing is shown in Table \ref{tab:dataset}. We further partitioned the gene set according to a causal partition defined in \citet{shah2025causal} using an initial undirected structure to obtain overlapping subsets of genes that retain consistency of causal discovery. For our work, we use the protein-protein interaction network in the STRING database as the initial undirected structure, and the maximum subset size in the partition of the structure was 2866 genes. A final DAG is obtained using the merge algorithm outlined in \citet{shah2025causal} which is based on a union of the subgraphs. We estimate a DAG at each dose rate $d \in D$ corresponding to a subset of $X_\text{TPM}$. We bootstrap \texttt{DAG-GNN} over 10 resamples of the data with replacement to produce a distribution of DAGs at each dose rate. Further, we derive a consensus graph over the distribution of DAGs, where we take the union over edges that co-occur in 50\% or more of the runs. To determine important genes at each dose rate, we take the nodes of the consensus graphs with at least one edge. In practice, many nodes end up being islands in the consensus graphs, which results in a gene set at each dose rate that is significantly smaller than the entire gene set $p=15,694$. For a description of the partitioning algorithm, training procedure and selected hyperparameters see Appendix \ref{appendix:training}.

\begin{table}
\caption{Dimensions of each $X_{\text{TPM}_d}$ for each dose rate after pre-processing.}
\label{tab:dataset}
  \centering
  \begin{tabular}{c c c}
     Dose Rate $d$ & \# Genes & \# Samples\\
    (mGy/hr) & $p$  &  $n$ \\
    \hline
        \midrule
     0.28 & 9262 & 36  \\
     0.38 & 9414  & 36 \\
     0.55 & 6274 & 36 \\
     6.66 &  8326 & 36 \\
     12.11 & 10603 & 36  \\
  \end{tabular}
  \end{table}
\subsection{Evaluation}
\subsubsection{Pathway Enrichment}
After important genes are identified, pathway enrichment is used to understand the functional roles these genes play in the exposed condition. We use the Python interface for \texttt{gProfiler} for pathway enrichment analysis of genes identified from differential expression analysis, supervised machine learning with random forest regression models, and causal discovery. \texttt{gProfiler} performs functional profiling of gene lists using various sources of biological evidence (Gene Ontology terms, biological pathways, regulatory DNA elements, human disease gene annotations, and protein-protein interaction networks). \texttt{gProfiler} uses Fisher's one-tailed test as the $p$-value measuring the randomness of the intersection between the query gene set and the pathway term. The higher the $-\text{log}_{10}p$-value the stronger the enrichment of that term is in the gene set. We provide a set of background genes (all genes in $X_{\text{TPM}}$) to ensure pathway enrichment $p$-values are comparable to each other across queries. We screen for pathways with \texttt{term\_size} $<$ 300 to filter out broad biological processes in the Gene Ontology that have thousands of genes.  


\subsubsection{Graph Structural Analysis}
We validate the edge set and graph topologies of the DAG distributions inferred at each dose rate. We perform TF enrichment to determine if nodes with high out-degree (hub nodes) overlap with known TF genes which code for proteins that regulate hundreds of other genes. We use TRRUST \citep{han2018trrust} and a ChIP-Seq network constructed with ENCODE \citep{davis2018encyclopedia} and Chip-Atlas \citep{zou2022chip} knowledge bases to identify human transcription factors and directed regulatory edges. Following the strategy of \cite{chevalley2025large} we use networks from a well studied  cell line HepG2 because the RPE1 cell line has not been as comprehensively characterized (both cell lines are epithelial). We measure the overlap of edges in the DAGs with known protein-protein interactions (STRING \citet{mering2003string} and CORUM \citet{giurgiu2019corum}), and regulatory edges (TRRUST, ChIP-seq network). We report the precision and F1-scores of the consensus graphs at each dose rate. 

\section{Results} 
\label{sec:results}
\begin{figure}[H]
    \centering
    \includegraphics[width=0.7\linewidth]{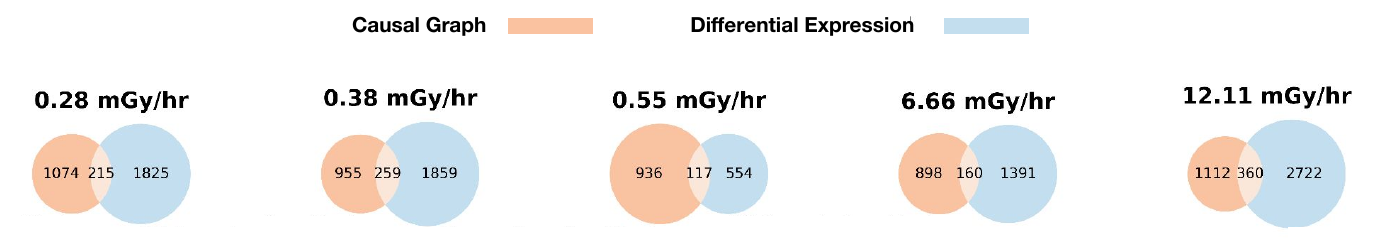}
    \caption{Venn diagrams showing gene overlaps between causal graphs and differential expression at each dose rate.}
    \label{fig:gene_sets}
\end{figure}

Differential expression analysis with \texttt{DESeq2} and causal discovery with \texttt{DAG-GNN} both identify gene sets at each dose rate. Gene sets and intersections are visualized as Venn diagrams in Fig. \ref{fig:gene_sets}. Differential expression identifies larger sets in general, with the exception of dose rate 0.55 mGy/hr. Gene sets between methods have small overlap, the highest at dose rate 12.11 mGy/hr -- these are also the largest gene sets for each method. Supervised machine learning with \texttt{RandomForestRegression} models yielded 68 genes inferred across all doses and have minimal overlap with causal graph and differential expression gene sets (Appendix Fig. \ref{fig:venn_rf}).

\begin{figure}[H]
    \centering
    \includegraphics[width=\linewidth]{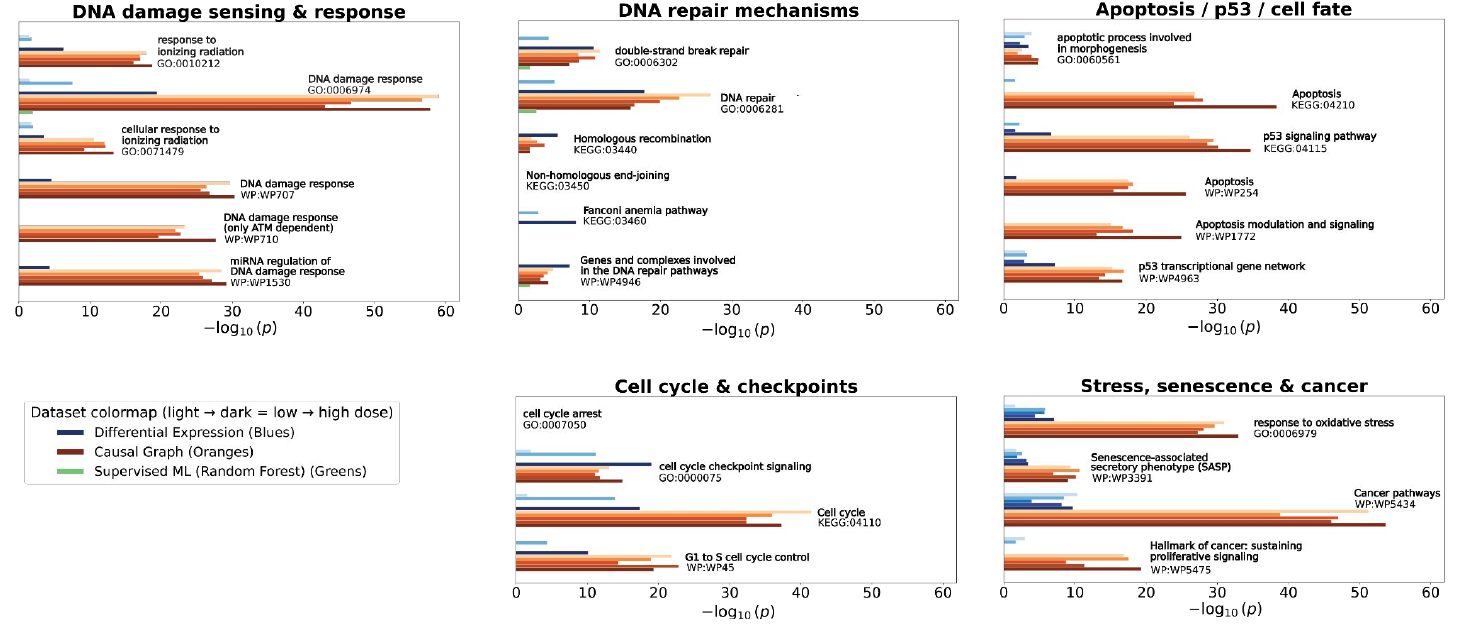}
    \caption{Pathway enrichment with \texttt{gProfiler} for manually curated radiation pathways across knowledge bases, categorized by radiation response type.} 
    \label{fig:pathway-enrichment-radiation}
\end{figure}

Pathway enrichment with \texttt{gProfiler} is shown in Fig. \ref{fig:pathway-enrichment-radiation} for a set of manually curated radiation response specific pathways from the Gene Ontology, KEGG, WikiPathways and Reactome. Causal graph gene sets have higher enrichment across radiation response pathways, indicating that causal gene sets better identify important genes for radiation response with smaller gene sets. Supervised machine learning did not enrich any radiation pathways with the exception of \texttt{Genes and complexes involved in DNA repair pathways}. Certain pathways such as \texttt{Non-homologous end-joining} and \texttt{Cell cycle arrest} were not enriched in any method. We do not see a significant dose-dependent enrichment signal; that is, higher dose rates do not correspond to specific pathways compared to lower dose rates. We validate our enrichment results against random genes from the background set and genes that have high correlation with the cumulative radiation (Appendix Fig. \ref{fig:sanity_check}).

\begin{figure}[htpb!]
    \centering
    \includegraphics[width=0.75\textwidth]{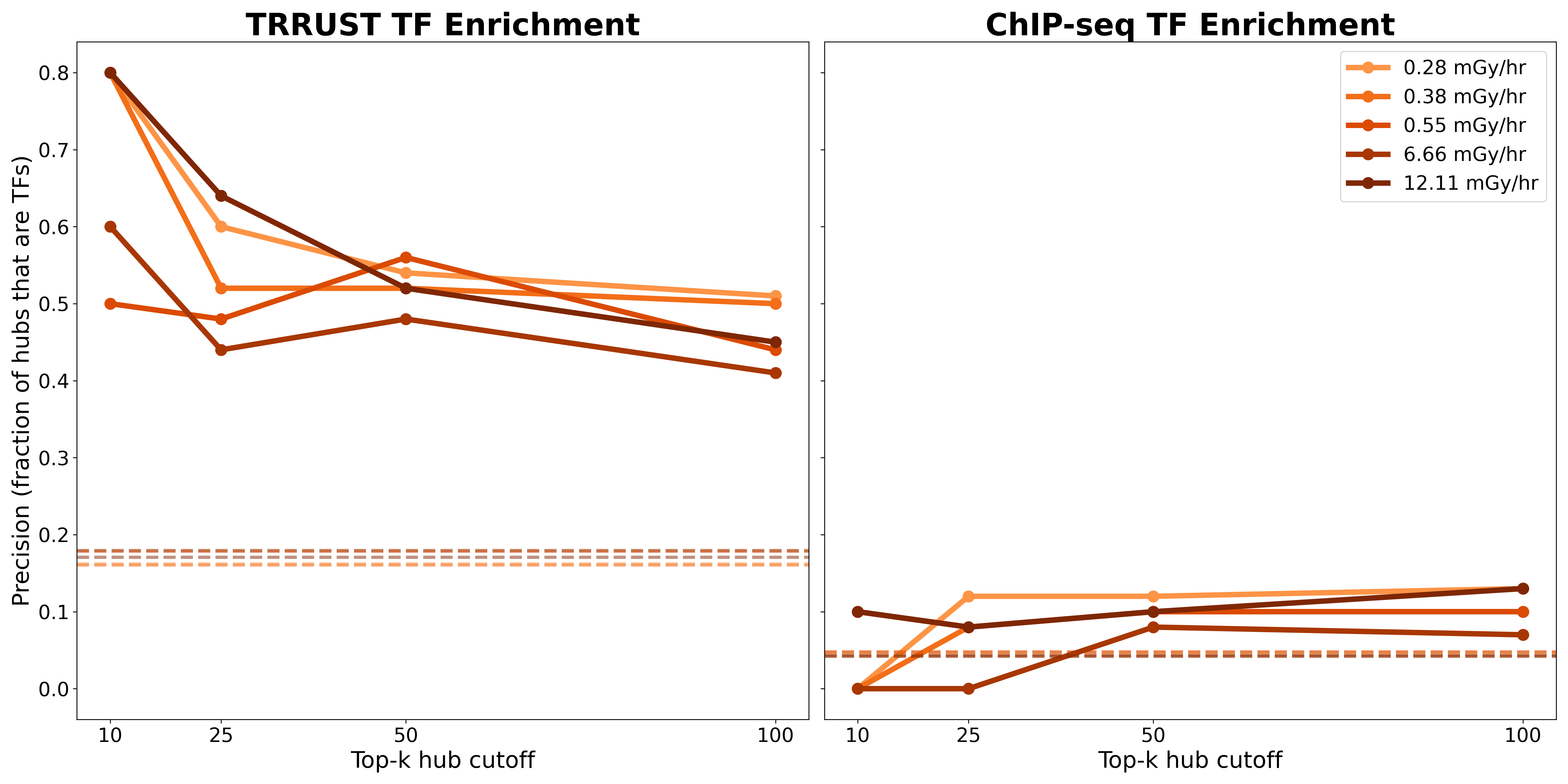}   
    \caption{Percent of top-k out-degree nodes (hubs) that are known transcription factor genes for high out-degree nodes by dose rate for TRRUST and ChIP-seq knowledge bases. Dashed lines represent the baseline fraction of known transcription factors are in the graph's node set.}
    \label{fig:hub_tf_enrichment}
\end{figure}

TF enrichment for each causal graph is shown in Figure \ref{fig:hub_tf_enrichment}. We see that for the top $k=10$ hub genes in the causal graph, up to 80\% are verified as genes that code for transcription factors in the TRRUST database. This percentage decreases as $k$ increases; however, enrichment is always above the baseline random assignment of transcription factors to genes (shown in dashed lines). TF enrichment compared to the ChIP-Seq network shows significantly less overlap; this may be because of the mismatch in cell types since there is no comprehensive network for RPE1 cell lines.

\begin{figure}[H]
    \centering

    \begin{minipage}[t]{0.60\textwidth}
        \centering
        \includegraphics[width=\linewidth]{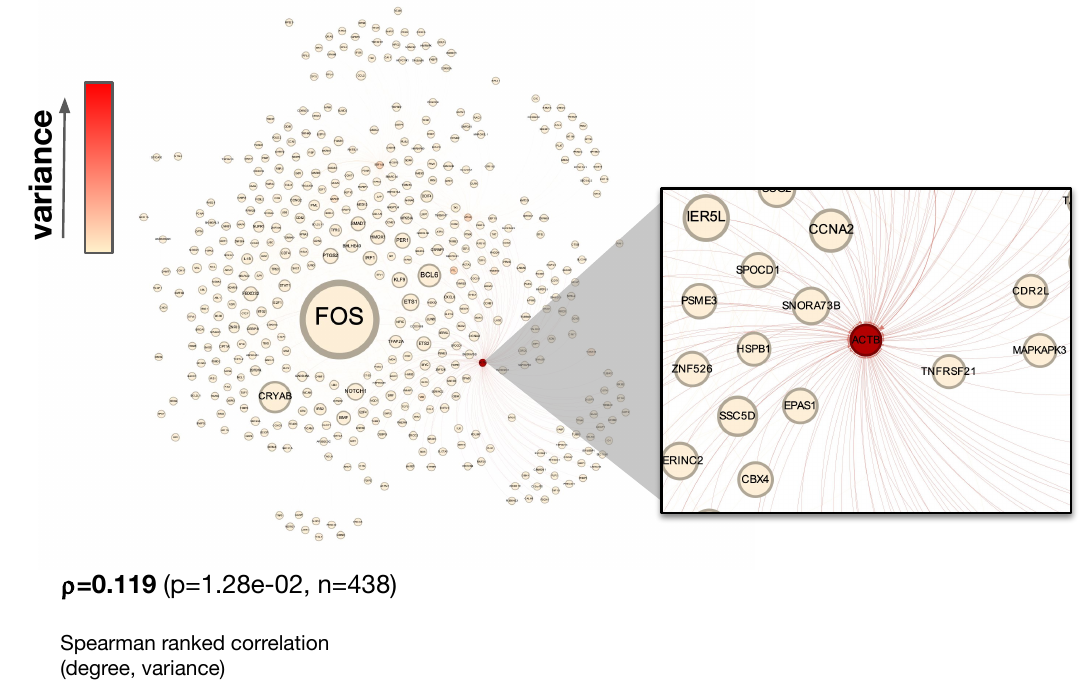}
    \end{minipage}
    \hfill
    \begin{minipage}[t]{0.35\textwidth}
        \centering
        \includegraphics[width=\linewidth]{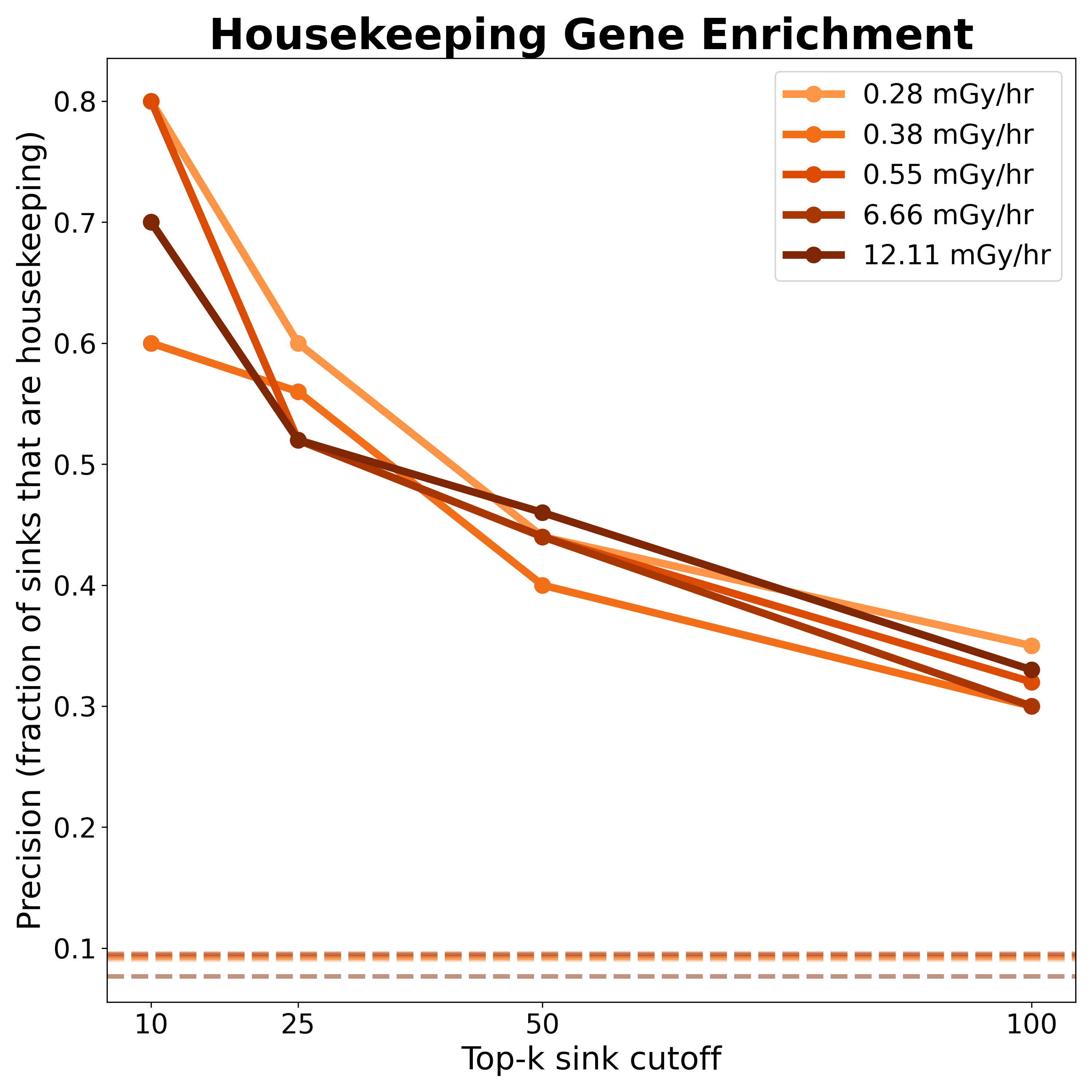}
    \end{minipage}

    \caption{(\textbf{Left}) An example causal graph at dose rate 6.66 mGy/hr, with nodes colored by variance. Several highly connected sink nodes correspond to housekeeping genes; ACTB is shown in the zoomed-in panel. (\textbf{Right}) Percent of top-$k$ in-degree nodes (sinks) that are housekeeping genes for each dose rate. Dashed lines represent the baseline fraction of housekeeping genes in the graph's gene set.}
    \label{fig:variance_housekeeping}
\end{figure}
To rule out the possibility that causal discovery simply assigns highly variant genes as hub nodes, we compute the Spearman ranked correlation between node degree and variance in $X_{\text{TPM}}$; see Fig. \ref{fig:variance_housekeeping} on the left. We find that the highest correlation from all dose rates is $\rho=0.119$. This means there is a small tendency for higher-degree genes in the causal graph to have higher expression variance, but causal discovery is not just recapitulating high-variance genes and instead is finding biologically meaningful structure. This is evidenced by high out-degree nodes mapping to known TFs which are master regulators, and high in-degree nodes map to known housekeeping genes (Fig. \ref{fig:variance_housekeeping} on the right) which are maintenance genes essential for cellular function. Housekeeping genes have also recently been shown to have high instability in stress conditions \citep{eisenberg2013human}, which includes radiation exposure \footnote{Housekeeping gene set from \citet{hsiao2001compendium}.}.


\begin{table}[htbp!]
\centering
\caption{Edge overlap between causal graphs and prior knowledge databases.
  TP = true positive number of edges found in the knowledge graph ;
  F1 = harmonic mean of precision and recall.
  For TRRUST and ChIP-seq, directed and reversed true positives are shown separately.}
\label{tab:edge_overlap}
\small
\setlength{\tabcolsep}{4pt}
\begin{tabular}{cc|ccc|cc|cc|ccc}
\toprule
 Dose & & \multicolumn{3}{c}{TRRUST} & \multicolumn{2}{c}{STRING} & \multicolumn{2}{c}{CORUM} & \multicolumn{3}{c}{ChIP-seq} \\
\cmidrule(lr){3-5}\cmidrule(lr){6-7}\cmidrule(lr){8-9}\cmidrule(lr){10-12}
(mGy/hr) & Edges & TP$_\text{dir}$ & TP$_\text{rev}$ & F1 & TP & F1 & TP & F1 & TP$_\text{dir}$ & TP$_\text{rev}$ & F1 \\
\midrule
\midrule
0.28 & 2777 & 24 & 12 & 0.012 & 143 & 0.026 & 54 & 0.020 & 111 & 107 & 0.006 \\
0.38 & 2610 & 17 & 7 & 0.009 & 130 & 0.025 & 47 & 0.019 & 112 & 80 & 0.006 \\
0.55 & 1248 & 3 & 4 & 0.003 & 52 & 0.016 & 26 & 0.019 & 54 & 16 & 0.004 \\
6.66 & 2263 & 16 & 8 & 0.010 & 113 & 0.031 & 34 & 0.019 & 88 & 82 & 0.006 \\
12.11 & 2417 & 18 & 13 & 0.009 & 142 & 0.025 & 36 & 0.015 & 107 & 89 & 0.004 \\
\bottomrule
\end{tabular}
\end{table}

In Table \ref{tab:edge_overlap}, we show the amount of edge overlap between causal graph at each dose rate and different knowledge graphs, including undirected protein-protein interaction graphs or complexes (STRING, CORUM) and directed regulatory graphs (TRRUST, ChIP-Seq). We see minimal overlap, with a maximum true positive edge count of 143, and maximum F1 score of 0.031. This is similar to results from \citet{chevalley2025large}, which used larger scale single-cell datasets for causal discovery. 

The set of \enquote{invariant} causal graph nodes is the intersection of nodes across all dose rates and results in 438 genes. We performed pathway enrichment on this invariant gene set, which we hypothesized as encoding fundamental radiation response relationships. The top 10 enriched pathways are shown in Table \ref{tab:invariant_top10_causal}. Note that unlike previous enrichment analyses, here we report the top pathways without explicit manual curation of radiation-specific pathways. Despite this, all top pathways correspond to well-known radiation response pathways \citep{sasaki2002dna, feinendegen2004responses}. We include the top 10 pathways for differential expression which have some overlap to radiation response (specifically inflammation pathways and tumor necrosis) but have larger $p$-values in Appendix Table \ref{tab:invariant_top10_de_invariant}.

\begin{table}[H]
\centering
\caption{Top 10 Enriched Pathways for  the Causal Invariant Gene Set ($n=438$)}
\label{tab:invariant_top10_causal}
\begin{tabular}{p{6cm} r r r}
\toprule
Pathway & $-\log_{10}(p)$ & Term Size & Intersection Size\\
\midrule
\midrule
Cell cycle \newline {\texttt{WP:WP179}} & 26.76 & 120 & 31 \\
\midrule
p53 signaling pathway \newline { \texttt{KEGG:04115}} & 26.48 & 74 & 26 \\
\midrule
Cell cycle \newline {\texttt{KEGG:04110}} & 25.94 & 157 & 33 \\
\midrule
miRNA regulation of DNA damage response \newline {\texttt{WP:WP1530}} & 23.34 & 73 & 24 \\
\midrule
signal transduction by p53 class mediator \newline {\texttt{GO:0072331}} & 23.02 & 164 & 31 \\
\midrule
Interleukin-4 and Interleukin-13 signaling \newline {\texttt{REAC:R-HSA-6785807}} & 22.70 & 111 & 27 \\
\midrule
DNA damage response \newline {\small \texttt{WP:WP707}} & 22.54 & 69 & 23 \\
\midrule
Integrated breast cancer pathway \newline {\texttt{WP:WP1984}} & 22.45 & 152 & 30 \\
\midrule
Cellular senescence \newline {\small \texttt{KEGG:04218}} & 22.43 & 155 & 30 \\
\midrule
Mitotic G1 phase and G1/S transition \newline {\texttt{REAC:R-HSA-453279}} & 21.65 & 148 & 29 \\
\midrule
\bottomrule
\end{tabular}
\end{table}

\begin{figure}[htpb]
    \centering
    \includegraphics[width=\textwidth]{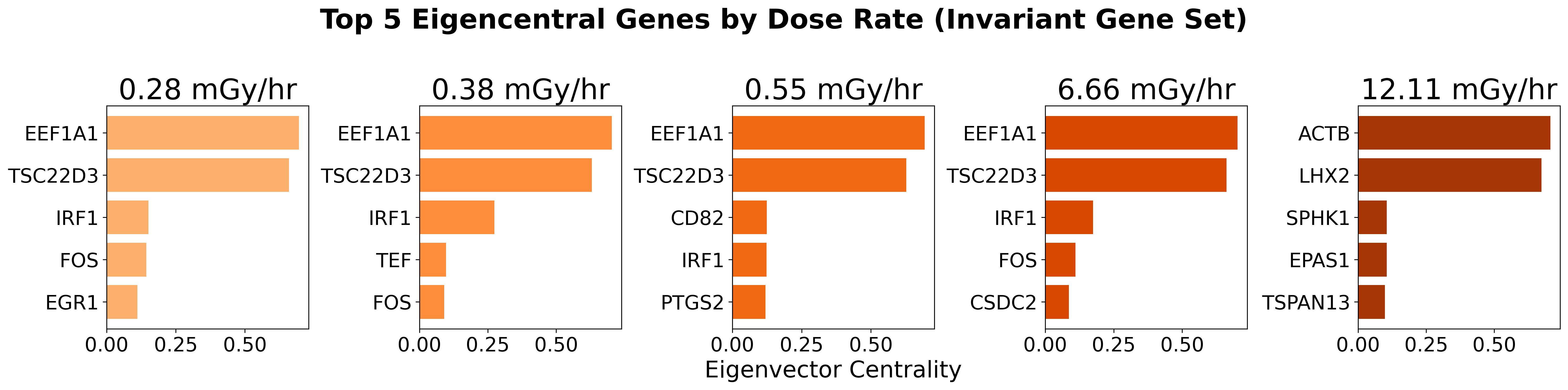}  
    \caption{The top 5 identified eigencentric genes for each dose-rate causal subgraph induced by the invariant set of genes.}
    \label{fig:eigencentrality}
\end{figure}
\begin{figure}[htpb!]
    \centering
    \includegraphics[width=\textwidth]{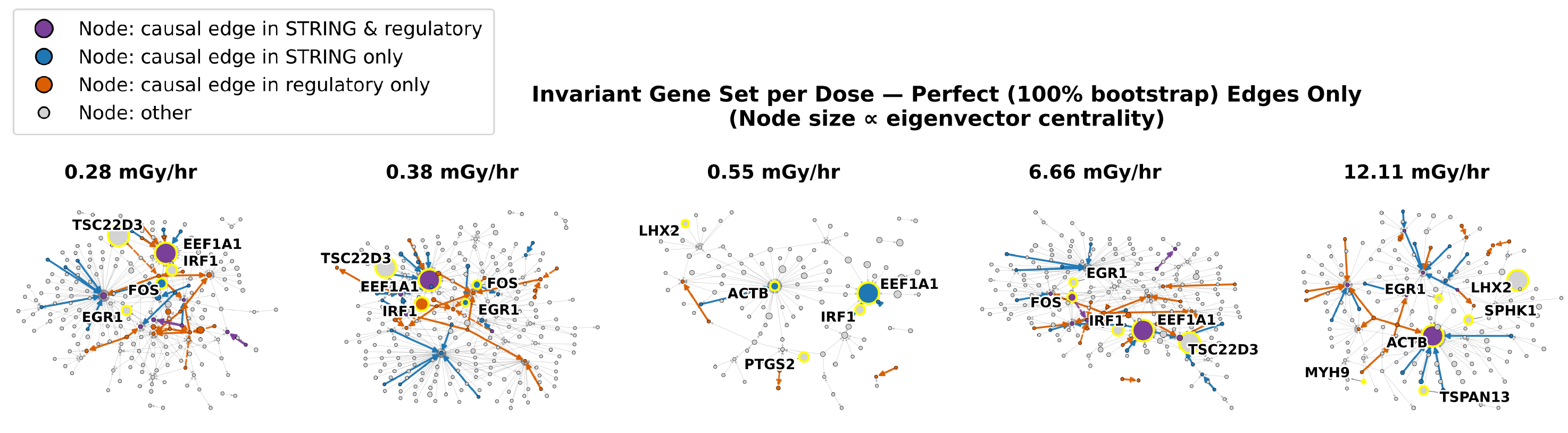}  
    \caption{The invariant subgraphs at each dose rate with only the \enquote{perfect} edges that co-occur across 100\% of the bootstrap distribution visualized. Annotated genes correspond to the top-5 eigencentric genes at each dose rate. Edges are highlighted according to true positives in knowledge bases where \enquote{regulatory} means either TRRUST or ChIP-seq edges.}
    \label{fig:invariant_subgraphs}
\end{figure}

We identify causal subgraphs over the invariant gene set at each dose rate and compute the top eigencentral genes (a measure of the node's influence in the subgraph) in Fig. \ref{fig:eigencentrality}. We find that top scoring genes correspond to transcription factors (FOS, TSC22D3, IRF1, EGR1, TEF, LHX2, EPAS1) and housekeeping genes (EEF1A1, ACTB), again validating the known structure regarding these genes. Notably, the highest dose-rate (12.11 mGy/hr) has different eigencentric genes suggesting a different biological structure compared to the lower dose rates. We additionally visualize a subset of edges in the subgraphs over the invariant gene set at each dose rate in Fig. \ref{fig:invariant_subgraphs} corresponding to \enquote{perfect} edges that co-occur across 100\% of the bootstrap distribution. These correspond to stable edges identified by \texttt{DAG-GNN} and, although only a handful are true positives in knowledge bases, these edges could be candidates for novel relationships with biological significance in radiation response. 

\begin{figure}[htpb!]
    \centering
    \begin{minipage}[t]{0.51\textwidth}
        \centering
        \includegraphics[width=\linewidth]{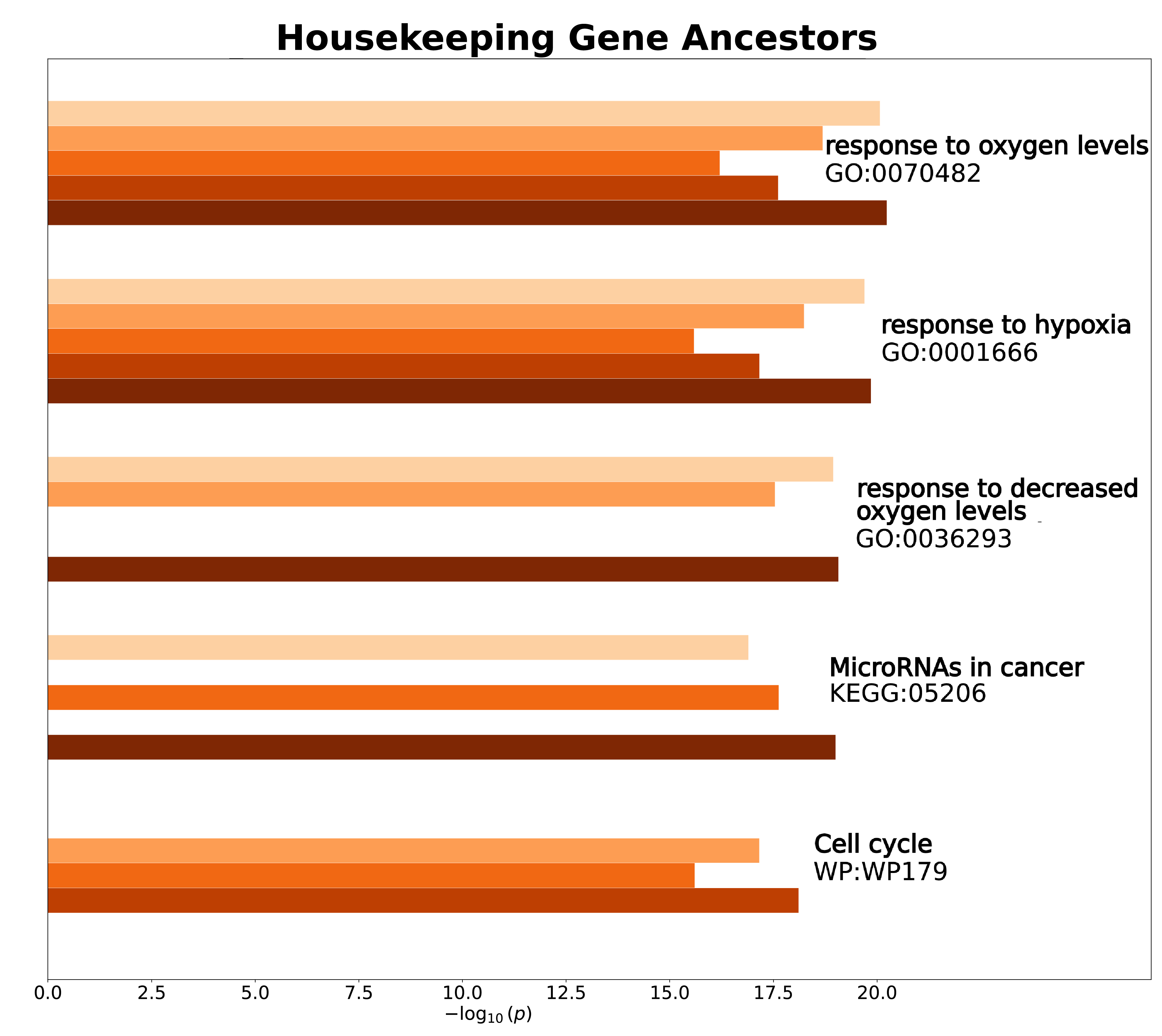}
        \label{fig:housekeeping_parents}
    \end{minipage}
    \hfill
    \begin{minipage}[t]{0.45\textwidth}
        \centering
        \includegraphics[width=\linewidth]{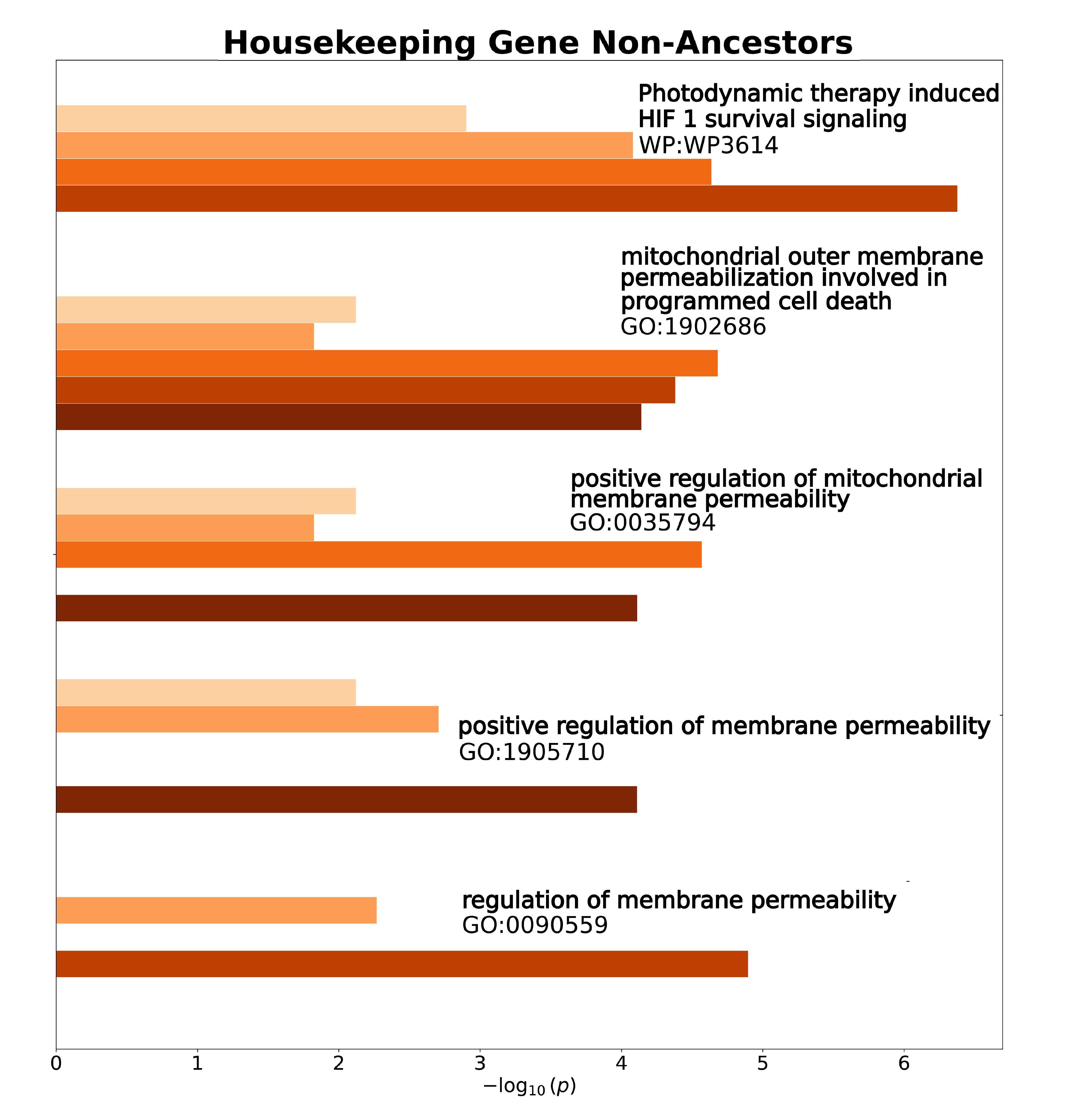}

        \label{fig:housekeeping_nonancestors}
    \end{minipage}

    \caption{(\textbf{Left}) Enrichment of housekeeping genes in the parent sets of the causal subgraphs over the invariant gene set. (\textbf{Right}) Enrichment of housekeeping genes among the non-ancestors in the causal subgraphs over the invariant gene set.}
    \label{fig:housekeeping}
\end{figure}

Housekeeping genes had a high overlap in the causal invariant gene set (14.4\%) compared to the differential expression invariant gene set (1.5\%). In Fig. \ref{fig:housekeeping}, we perform pathway enrichment on the ancestors compared to non-ancestors of housekeeping genes. We find that the enrichment profile between these two groups is distinct: the ancestor set overlaps with response to oxidative stress (ROS) pathways, and the non-ancestor set overlaps with membrane permeability pathways. This suggests a branching effect between stress (ROS) and apoptosis (membrane permeability is an indicator of triggered cell death). This branching effect is well documented in the literature \citep{fulda2010cellular}.

The results reported in this section use the consensus causal graphs over \texttt{DAG-GNN} bootstrap runs, which filter for edges that co-occur in more than 50\% of the bootstrap runs. Analyses for each DAG in the bootstrap is in Appendix \ref{appendix:bootstrap}.
\section{Discussion} 
\label{sec:discussion}

We present a previously unexplored application of causal discovery with transcriptomics data to identify important genes related to perturbation response. Typically differential expression analysis is performed to identify important genes by comparing a control group to perturbed groups. These genes are then used for pathway enrichment to identify possible biomarkers of the perturbation condition. We demonstrate that differential expression analysis may potentially miss important genes related to perturbation response. By jointly modeling the perturbation variable and the multivariate gene expression distribution we show that causal discovery has a higher hit rate of finding perturbation response related gene sets, using radiation response as a use case. Furthermore, the learned DAG identifies structural signatures such as transcription factors and housekeeping genes and hub and sink nodes respectively. We find that causal subgraphs induced by a core set of invariant genes are enriched for perturbation-specific pathways and appear to encode branching between apoptosis and stress response pathways. This observation aligns with invariant causal prediction \citep{peters2016causal}, which proposes that causal relationships exhibit stability across environments, here corresponding to different dose rates. To infer novel pathways with causal discovery, one can identify edges that co-occur across the entire bootstrap distribution corresponding to stable edges with potential biological significance. These novel pathways should then be validated with experimentation, for example, with CRISPR knockouts of upstream genes.

Applications of causal discovery have focused on exact recovery of causal edges; this is a notoriously difficult task when we consider the numerous ways in which most realistic datasets violate assumptions needed for identifiability. Research currently focuses on collecting large-scale perturbational datasets to improve identification of the true causal graph. However, there is a need for interpretable methods for biomarker and mechanism discovery in low-sample settings (e.g., small patient cohorts, rare diseases, or settings in which high throughput data is difficult to collect). In the presence of strong upstream perturbations (e.g., environmental stress or drug dosage), we propose that causal discovery should be considered as part of the bioinformatics toolbox as existing methods may be suboptimal in this setting.


\paragraph{Limitations}
Our focus on radiation response may limit the generalizability of our findings. Future work will extend this analysis to additional transcriptomic datasets to evaluate the broader applicability of causal discovery for modeling perturbation response. We note that causal discovery incurs substantially higher computational cost than differential expression analysis and supervised machine learning (Appendix \ref{appendix:training}); this limits large-scale statistical evaluation. As a result, our current analysis is primarily qualitative, relying on consensus graphs across bootstrap runs. A more rigorous quantification of these findings through statistical analysis over distributions of consensus graphs is an important direction for future work. Finally, we acknowledge that pathway enrichment is biased toward well-studied genes and can be prone to over-interpretation.

\bibliography{radbio}

@article{jantre2026interpretable,
  title={Interpretable transcriptome-to-phenotype modeling of cell-painting nuclear morphology features from RNA-seq under low-dose radiation exposure},
  author={Jantre, Sanket and Chopra, Kriti and Zhao, Guang and Cucinell, Clark and Weinberg, Rebecca and Forrester, Sara and Brettin, Thomas and Urban, Nathan M and Qian, Xiaoning and Yoon, Byung-Jun},
  journal={bioRxiv},
  pages={2026--02},
  year={2026},
  publisher={Cold Spring Harbor Laboratory}
}

@article{wenric2018using,
  title={Using supervised learning methods for gene selection in RNA-Seq case-control studies},
  author={Wenric, Stephane and Shemirani, Ruhollah},
  journal={Frontiers in genetics},
  volume={9},
  pages={364621},
  year={2018},
  publisher={Frontiers}
}

@article{marbach2012wisdom,
  title={Wisdom of crowds for robust gene network inference},
  author={Marbach, Daniel and Costello, James C and K{\"u}ffner, Robert and Vega, Nicole M and Prill, Robert J and Camacho, Diogo M and Allison, Kyle R and Kellis, Manolis and Collins, James J and others},
  journal={Nature methods},
  volume={9},
  number={8},
  pages={796--804},
  year={2012},
  publisher={Nature Publishing Group US New York}
}

@article{chen2022graph,
  title={Graph attention network for link prediction of gene regulations from single-cell RNA-sequencing data},
  author={Chen, Guangyi and Liu, Zhi-Ping},
  journal={Bioinformatics},
  volume={38},
  number={19},
  pages={4522--4529},
  year={2022},
  publisher={Oxford University Press}
}

@inproceedings{yu2019dag,
  title={DAG-GNN: DAG structure learning with graph neural networks},
  author={Yu, Yue and Chen, Jie and Gao, Tian and Yu, Mo},
  booktitle={International conference on machine learning},
  pages={7154--7163},
  year={2019},
  organization={PMLR}
}

@article{raudvere2019g,
  title={g: Profiler: a web server for functional enrichment analysis and conversions of gene lists (2019 update)},
  author={Raudvere, Uku and Kolberg, Liis and Kuzmin, Ivan and Arak, Tambet and Adler, Priit and Peterson, Hedi and Vilo, Jaak},
  journal={Nucleic acids research},
  volume={47},
  number={W1},
  pages={W191--W198},
  year={2019},
  publisher={Oxford University Press}
}

@article{love2014moderated,
  title={Moderated estimation of fold change and dispersion for RNA-seq data with DESeq2},
  author={Love, Michael I and Huber, Wolfgang and Anders, Simon},
  journal={Genome biology},
  volume={15},
  number={12},
  pages={550},
  year={2014},
  publisher={Springer}
}

@article{lu2005hotelling,
  title={Hotelling's T 2 multivariate profiling for detecting differential expression in microarrays},
  author={Lu, Yan and Liu, Peng-Yuan and Xiao, Peng and Deng, Hong-Wen},
  journal={Bioinformatics},
  volume={21},
  number={14},
  pages={3105--3113},
  year={2005},
  publisher={Oxford University Press}
}

@article{klebanov2007multivariate,
  title={A multivariate extension of the gene set enrichment analysis},
  author={Klebanov, Lev and Glazko, Galina and Salzman, Peter and Yakovlev, Andrei and Xiao, Yuanhui},
  journal={Journal of bioinformatics and computational biology},
  volume={5},
  number={05},
  pages={1139--1153},
  year={2007},
  publisher={World Scientific}
}

@article{glazko2009unite,
  title={Unite and conquer: univariate and multivariate approaches for finding differentially expressed gene sets},
  author={Glazko, Galina V and Emmert-Streib, Frank},
  journal={Bioinformatics},
  volume={25},
  number={18},
  pages={2348--2354},
  year={2009},
  publisher={Oxford University Press}
}

@article{feng2023gene,
  title={Gene regulatory network inference based on causal discovery integrating with graph neural network},
  author={Feng, Ke and Jiang, Hongyang and Yin, Chaoyi and Sun, Huiyan},
  journal={Quantitative Biology},
  volume={11},
  number={4},
  pages={434--450},
  year={2023},
  publisher={Wiley Online Library}
}

@article{chevalley2025large,
  title={A large-scale benchmark for network inference from single-cell perturbation data},
  author={Chevalley, Mathieu and Roohani, Yusuf H and Mehrjou, Arash and Leskovec, Jure and Schwab, Patrick},
  journal={Communications Biology},
  volume={8},
  number={1},
  pages={412},
  year={2025},
  publisher={Nature Publishing Group UK London}
}

@article{dixit2016perturb,
  title={Perturb-Seq: dissecting molecular circuits with scalable single-cell RNA profiling of pooled genetic screens},
  author={Dixit, Atray and Parnas, Oren and Li, Biyu and Chen, Jenny and Fulco, Charles P and Jerby-Arnon, Livnat and Marjanovic, Nemanja D and Dionne, Danielle and Burks, Tyler and Raychowdhury, Raktima and others},
  journal={cell},
  volume={167},
  number={7},
  pages={1853--1866},
  year={2016},
  publisher={Elsevier}
}

@article{zheng2018dags,
  title={Dags with no tears: Continuous optimization for structure learning},
  author={Zheng, Xun and Aragam, Bryon and Ravikumar, Pradeep K and Xing, Eric P},
  journal={Advances in neural information processing systems},
  volume={31},
  year={2018}
}

@article{pratapa2020benchmarking,
  title={Benchmarking algorithms for gene regulatory network inference from single-cell transcriptomic data},
  author={Pratapa, Aditya and Jalihal, Amogh P and Law, Jeffrey N and Bharadwaj, Aditya and Murali, andT M},
  journal={Nature methods},
  volume={17},
  number={2},
  pages={147--154},
  year={2020},
  publisher={Nature Publishing Group US New York}
}

@article{shimizu2010radiation,
  title={Radiation exposure and circulatory disease risk: Hiroshima and Nagasaki atomic bomb survivor data, 1950-2003},
  author={Shimizu, Yukiko and Kodama, Kazunori and Nishi, Nobuo and Kasagi, Fumiyoshi and Suyama, Akihiko and Soda, Midori and Grant, Eric J and Sugiyama, Hiromi and Sakata, Ritsu and Moriwaki, Hiroko and others},
  journal={Bmj},
  volume={340},
  year={2010},
  publisher={British Medical Journal Publishing Group}
}

@article{verheij2000radiation,
  title={Radiation-induced apoptosis},
  author={Verheij, Marcel and Bartelink, Harry},
  journal={Cell and tissue research},
  volume={301},
  number={1},
  pages={133--142},
  year={2000},
  publisher={Springer}
}

@article{rosati2024differential,
  title={Differential gene expression analysis pipelines and bioinformatic tools for the identification of specific biomarkers: A review},
  author={Rosati, Diletta and Palmieri, Maria and Brunelli, Giulia and Morrione, Andrea and Iannelli, Francesco and Frullanti, Elisa and Giordano, Antonio},
  journal={Computational and structural biotechnology journal},
  volume={23},
  pages={1154--1168},
  year={2024},
  publisher={Elsevier}
}

@article{brouillard2024landscape,
  title={The landscape of causal discovery data: Grounding causal discovery in real-world applications},
  author={Brouillard, Philippe and Squires, Chandler and Wahl, Jonas and Kording, Konrad P and Sachs, Karen and Drouin, Alexandre and Sridhar, Dhanya},
  journal={arXiv preprint arXiv:2412.01953},
  year={2024}
}

@article{
shah2025causal,
title={Causal Discovery over High-Dimensional Structured Hypothesis Spaces with  Causal Graph Partitioning},
author={Ashka Shah and Adela Frances DePavia and Nathaniel C Hudson and Ian Foster and Rick Stevens},
journal={Transactions on Machine Learning Research},
issn={2835-8856},
year={2025},
url={https://openreview.net/forum?id=FecsgPCOHk},
note={}
}

@article{trapnell2009tophat,
  title={TopHat: discovering splice junctions with RNA-Seq},
  author={Trapnell, Cole and Pachter, Lior and Salzberg, Steven L},
  journal={Bioinformatics},
  volume={25},
  number={9},
  pages={1105--1111},
  year={2009},
  publisher={Oxford University Press}
}

@article{moerman2019grnboost2,
  title={GRNBoost2 and Arboreto: efficient and scalable inference of gene regulatory networks},
  author={Moerman, Thomas and Aibar Santos, Sara and Bravo Gonz{\'a}lez-Blas, Carmen and Simm, Jaak and Moreau, Yves and Aerts, Jan and Aerts, Stein},
  journal={Bioinformatics},
  volume={35},
  number={12},
  pages={2159--2161},
  year={2019},
  publisher={Oxford University Press}
}

@article{gene2019gene,
  title={The gene ontology resource: 20 years and still GOing strong},
  author={Gene Ontology Consortium},
  journal={Nucleic acids research},
  volume={47},
  number={D1},
  pages={D330--D338},
  year={2019},
  publisher={Oxford University Press}
}

@inproceedings{kanehisa2002kegg,
  title={The KEGG database},
  author={Kanehisa, Minoru},
  booktitle={‘In silico’simulation of biological processes: Novartis Foundation Symposium 247},
  volume={247},
  pages={91--103},
  year={2002},
  organization={Wiley Online Library}
}

@article{milacic2024reactome,
  title={The reactome pathway knowledgebase 2024},
  author={Milacic, Marija and Beavers, Deidre and Conley, Patrick and Gong, Chuqiao and Gillespie, Marc and Griss, Johannes and Haw, Robin and Jassal, Bijay and Matthews, Lisa and May, Bruce and others},
  journal={Nucleic acids research},
  volume={52},
  number={D1},
  pages={D672--D678},
  year={2024},
  publisher={Oxford University Press}
}

@article{agrawal2024wikipathways,
  title={WikiPathways 2024: next generation pathway database},
  author={Agrawal, Ayushi and Balc{\i}, Hasan and Hanspers, Kristina and Coort, Susan L and Martens, Marvin and Slenter, Denise N and Ehrhart, Friederike and Digles, Daniela and Waagmeester, Andra and Wassink, Isabel and others},
  journal={Nucleic acids research},
  volume={52},
  number={D1},
  pages={D679--D689},
  year={2024},
  publisher={Oxford University Press}
}

@article{han2018trrust,
  title={TRRUST v2: an expanded reference database of human and mouse transcriptional regulatory interactions},
  author={Han, Heonjong and Cho, Jae-Won and Lee, Sangyoung and Yun, Ayoung and Kim, Hyojin and Bae, Dasom and Yang, Sunmo and Kim, Chan Yeong and Lee, Muyoung and Kim, Eunbeen and others},
  journal={Nucleic acids research},
  volume={46},
  number={D1},
  pages={D380--D386},
  year={2018},
  publisher={Oxford University Press}
}

@article{zou2022chip,
  title={ChIP-Atlas 2021 update: a data-mining suite for exploring epigenomic landscapes by fully integrating ChIP-seq, ATAC-seq and Bisulfite-seq data},
  author={Zou, Zhaonan and Ohta, Tazro and Miura, Fumihito and Oki, Shinya},
  journal={Nucleic acids research},
  volume={50},
  number={W1},
  pages={W175--W182},
  year={2022},
  publisher={Oxford University Press}
}

@article{davis2018encyclopedia,
  title={The Encyclopedia of DNA elements (ENCODE): data portal update},
  author={Davis, Carrie A and Hitz, Benjamin C and Sloan, Cricket A and Chan, Esther T and Davidson, Jean M and Gabdank, Idan and Hilton, Jason A and Jain, Kriti and Baymuradov, Ulugbek K and Narayanan, Aditi K and others},
  journal={Nucleic acids research},
  volume={46},
  number={D1},
  pages={D794--D801},
  year={2018},
  publisher={Oxford University Press}
}

@article{giurgiu2019corum,
  title={CORUM: the comprehensive resource of mammalian protein complexes—2019},
  author={Giurgiu, Madalina and Reinhard, Julian and Brauner, Barbara and Dunger-Kaltenbach, Irmtraud and Fobo, Gisela and Frishman, Goar and Montrone, Corinna and Ruepp, Andreas},
  journal={Nucleic acids research},
  volume={47},
  number={D1},
  pages={D559--D563},
  year={2019},
  publisher={Oxford University Press}
}

@article{mering2003string,
  title={STRING: a database of predicted functional associations between proteins},
  author={Mering, Christian von and Huynen, Martijn and Jaeggi, Daniel and Schmidt, Steffen and Bork, Peer and Snel, Berend},
  journal={Nucleic acids research},
  volume={31},
  number={1},
  pages={258--261},
  year={2003},
  publisher={Oxford University Press}
}

@article{eisenberg2013human,
  title={Human housekeeping genes, revisited},
  author={Eisenberg, Eli and Levanon, Erez Y},
  journal={TRENDS in Genetics},
  volume={29},
  number={10},
  pages={569--574},
  year={2013},
  publisher={Elsevier}
}

@article{fulda2010cellular,
  title={Cellular stress responses: cell survival and cell death},
  author={Fulda, Simone and Gorman, Adrienne M and Hori, Osamu and Samali, Afshin},
  journal={International journal of cell biology},
  volume={2010},
  number={1},
  pages={214074},
  year={2010},
  publisher={Wiley Online Library}
}

@article{peters2016causal,
  title={Causal inference by using invariant prediction: identification and confidence intervals},
  author={Peters, Jonas and B{\"u}hlmann, Peter and Meinshausen, Nicolai},
  journal={Journal of the Royal Statistical Society Series B: Statistical Methodology},
  volume={78},
  number={5},
  pages={947--1012},
  year={2016},
  publisher={Oxford University Press}
}

@article{hsiao2001compendium,
  title={A compendium of gene expression in normal human tissues},
  author={Hsiao, Li-Li and Dangond, Fernando and Yoshida, Takumi and Hong, Robert and Jensen, Roderick V and Misra, Jatin and Dillon, William and Lee, Kailin F and Clark, Kathryn E and Haverty, Peter and others},
  journal={Physiological genomics},
  volume={7},
  number={2},
  pages={97--104},
  year={2001},
  publisher={American Physiological Society}
}

@article{feinendegen2004responses,
  title={Responses to low doses of ionizing radiation in biological systems},
  author={Feinendegen, Ludwig E and Pollycove, Myron and Sondhaus, Charles A},
  journal={Nonlinearity in biology, toxicology, medicine},
  volume={2},
  number={3},
  pages={15401420490507431},
  year={2004},
  publisher={SAGE Publications Sage CA: Los Angeles, CA}
}

@article{sasaki2002dna,
  title={DNA damage response pathway in radioadaptive response},
  author={Sasaki, Masao S and Ejima, Yosuke and Tachibana, Akira and Yamada, Toshiko and Ishizaki, Kanji and Shimizu, Takashi and Nomura, Taisei},
  journal={Mutation Research/Fundamental and Molecular Mechanisms of Mutagenesis},
  volume={504},
  number={1-2},
  pages={101--118},
  year={2002},
  publisher={Elsevier}
}
\newpage
\appendix
\section{\texttt{DAG-GNN} training at Scale}
\label{appendix:training}
\texttt{DAG-GNN} is trained using a nested optimization loop where an inner loop updates the neural network parameters and adjacency matrix to minimize the evidence lower bound (ELBO) loss, while an outer loop enforces the DAG constraint using an augmented Lagrangian. A trainable adjacency matrix is represented by $A \in \mathbb{R}^{p\times p}$. The encoder is $z= (I-A^T)x$ for input $x \in \mathbb{R}^{n \times p}$, where $I$ is the identity matrix. The decoder is  $(I-A^T)^{-1}z$. We set the  embedding dimension to 64. We trained \texttt{DAG-GNN} for 300 epochs using Adam with a learning rate of 3e-3 (\texttt{gamma}=1.0, \texttt{lr\_decay}= 200). We use $n$ for the batch size, because we do not have sufficient samples for stable mini-batching. The maximum number of iterations for the Lagrangian optimization was set to 100. The threshold for the adjacency matrix $A$ is set to 0.3, which is the default value. 
 
\texttt{DAG-GNN} training is time and memory intensive. To learn an adjacency matrix over $p=15,694$ would be intractable.  We partitioned the gene set according to a causal partition \citep{shah2025causal} of the protein-protein interaction network from the STRING database, then merged the graphs by taking the union of edges. Training times depend on the gene subset size, which vary by dose rate. The average training time for \texttt{DAG-GNN} on one subset of data across bootstraps was 23.3 min. Training curves for one example subset are shown in Fig. \ref{fig:training}. Note that loss spikes occur when the augmented Lagrangian penalty $\rho$ is increased in the outer loop, abruptly strengthening the acyclicity constraint and perturbing the learned graph. This is expected behavior as the model rebalances reconstruction accuracy with enforcing a valid DAG structure. Models were trained on a NVIDIA Tesla V100-SXM2-32GB GPU. 

\begin{figure}[htpb!]
    \centering
    \begin{minipage}[t]{0.51\textwidth}
        \centering
        \includegraphics[width=\linewidth]{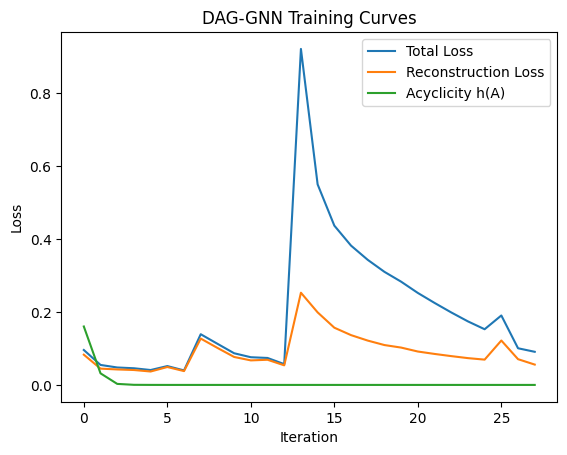}
        \label{fig:edge_conv}
    \end{minipage}
    \hfill
    \begin{minipage}[t]{0.45\textwidth}
        \centering
        \includegraphics[width=\linewidth]{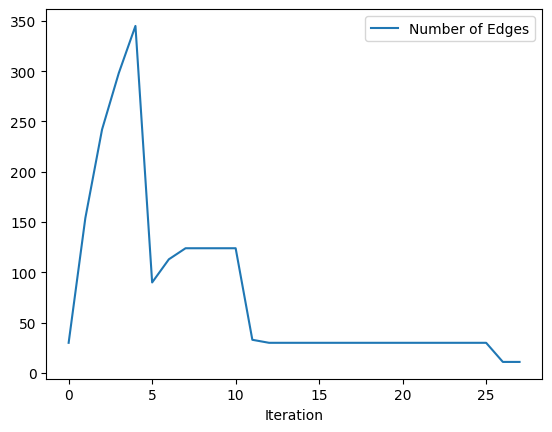}
        \label{fig:edge_cov}
    \end{minipage}

    \caption{(\textbf{Left}) Loss curves and acyclicity constraint $h(A)$ which converges to 0 to ensure a DAG. (\textbf{Right}) The number of edges in the learned adjacency matrix at each iteration using the default threshold 0.3.}
    \label{fig:training_edge_cov}
\end{figure}

\section{Linear Regression Baseline}
\label{appendix:linear}
\begin{figure}[H]
    \centering
    \includegraphics[width=\linewidth]{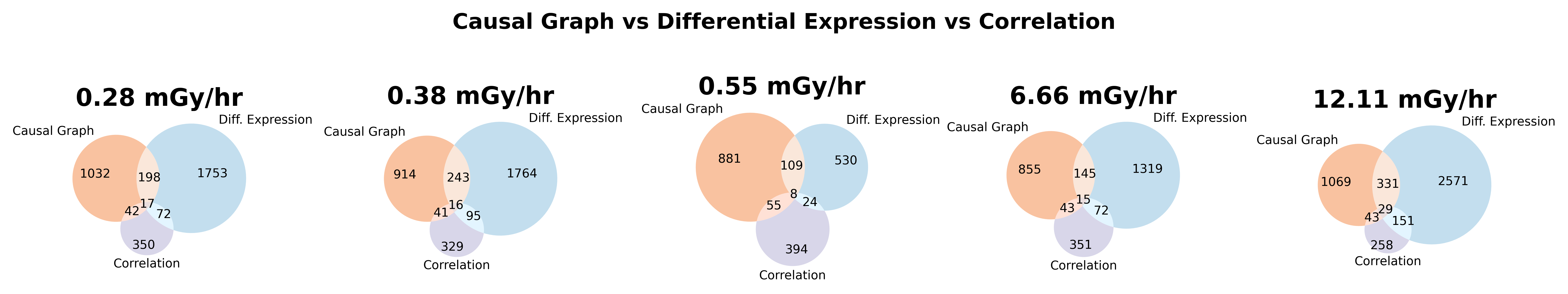}
    \caption{Intersections with the Correlation baseline algorithm (based on Linear Regression). }
    \label{fig:venn_corr}
\end{figure}

We were concerned  that causal gene sets might just be identifying genes that have strong correlative structure with the radiation vector, especially given a small sample size. We constructed a baseline algorithm based on linear regression of (week, dose\_rate) labels onto each gene. Each model's $R^2$ measures how much variance in that gene's expression is explained by dose rate and week jointly. A gene with a high multiple correlation $R^2$ is one whose expression varies strongly as a linear function of dose rate and/or time.  This captures genes that respond to radiation in a dose- or time-dependent manner — but purely through univariate linear association with the experimental conditions, without any causal graph structure or model-based feature importance. It serves as a \enquote{correlative baseline} against which the causal graph gene selections are compared. Similar to the supervised machine learning framework, we choose the 1000 genes at each fold with the top $R^2$ scores. Then we select genes that are stable in more than 50\% of folds. This results in 481 genes and is visualized as a \enquote{Correlation} baseline in Fig. \ref{fig:venn_corr}. We also perform pathway enrichment of this gene set and report the top 10 pathways in Table \ref{tab:top10_corr}. We see enrichment of radiation pathways as well, but a much higher enrichment for the invariant causal gene set (Table \ref{tab:invariant_top10_causal} which also has a smaller gene set size.
\begin{table}
\caption{Top 10 Enriched Pathways — Correlation Gene Set}
\label{tab:top10_corr}
\centering
\begin{tabular}{p{6cm} r r r}
\toprule
Pathway & $-\log_{10}(p)$ & Term Size & Intersection \\
\midrule
\midrule
Cell Cycle Checkpoints \newline {\small \texttt{REAC:R-HSA-69620}} & 8.28 & 289 & 24 \\
\hline
Regulation of TP53 Activity through Phosphorylation \newline {\small \texttt{REAC:R-HSA-6804756}} & 7.70 & 92 & 14 \\
\hline
Regulation of TP53 Activity \newline {\small \texttt{REAC:R-HSA-5633007}} & 7.18 & 160 & 17 \\
\hline
G1 to S cell cycle control \newline {\small \texttt{WP:WP45}} & 7.15 & 64 & 12 \\
\hline
Cell cycle \newline {\small \texttt{WP:WP179}} & 6.96 & 120 & 15 \\
\hline
Mitotic G1 phase and G1/S transition \newline {\small \texttt{REAC:R-HSA-453279}} & 6.82 & 148 & 16 \\
\hline
Retinoblastoma gene in cancer \newline {\small \texttt{WP:WP2446}} & 6.79 & 89 & 13 \\
\hline
Cell cycle \newline {\small \texttt{KEGG:04110}} & 6.30 & 157 & 16 \\
\hline
DNA damage response \newline \texttt{WP:WP707} & 6.02 & 69 & 11 \\
\hline
DNA replication \newline {\small \texttt{GO:0006260}} & 5.84 & 279 & 20 \\
\bottomrule
\end{tabular}
\end{table}

\section{Additional Gene Set Intersections}
\label{appendix:gene_sets}
\begin{figure}[H]
    \centering
    \includegraphics[width=\linewidth]{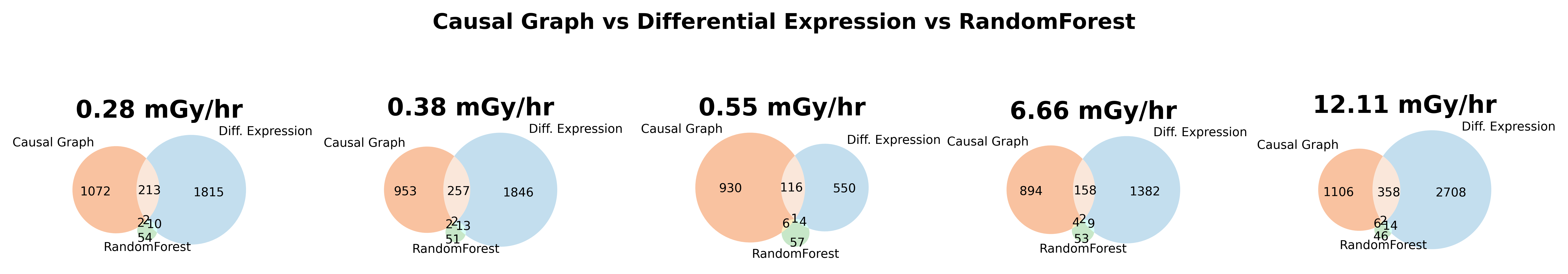}
    \caption{Intersections with the Random Forest gene set which is dose invariant.}
    \label{fig:venn_rf}
\end{figure}

\begin{figure}[htbp!]
    \centering
    \includegraphics[width=0.8\linewidth]{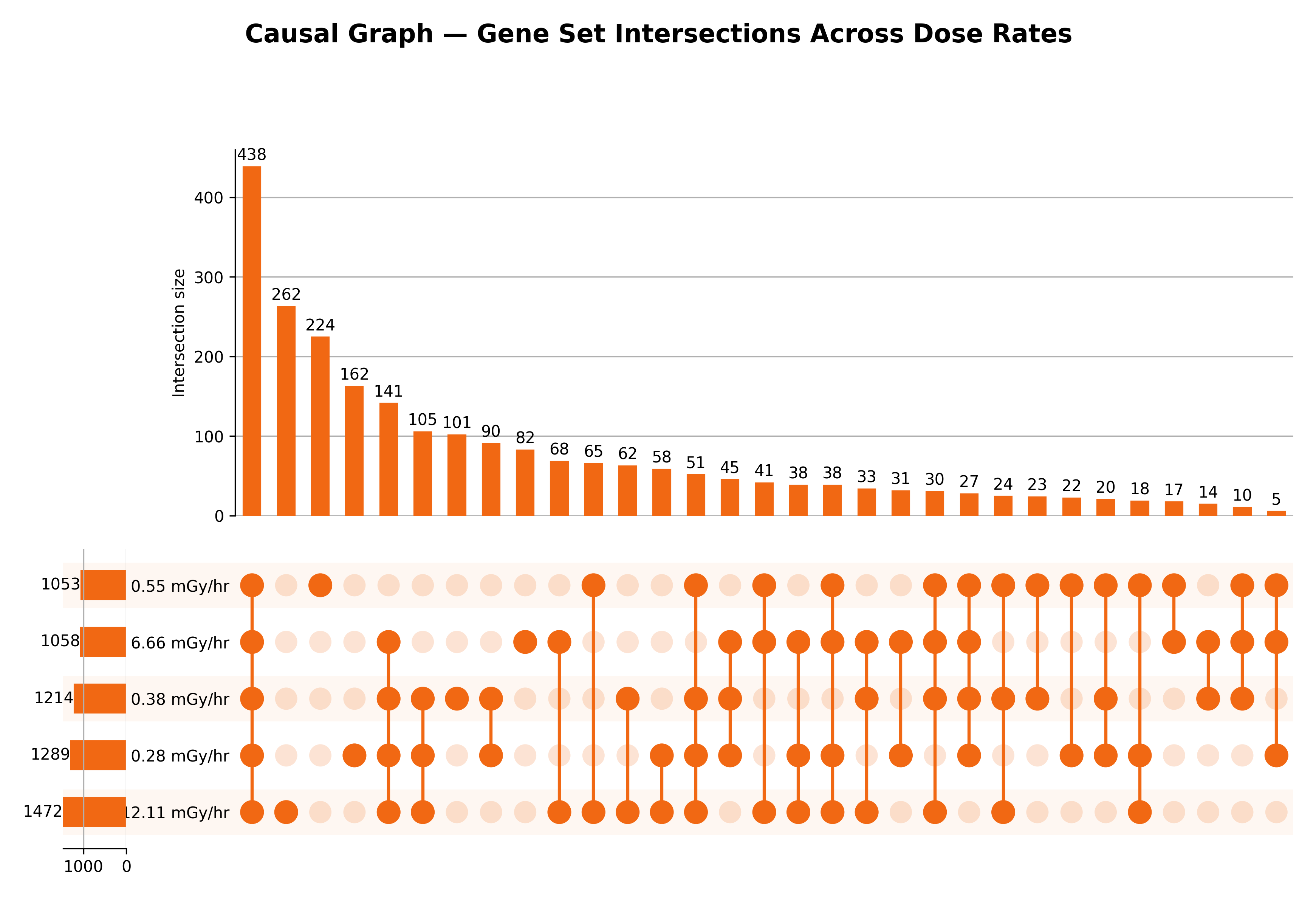}
    \caption{Intersections of causal graph gene sets across each dose rate. There is an \enquote{invariant} set of 438 genes which is the intersection across all dose rates.}
    \label{fig:upset_causal}
\end{figure}

\begin{figure}[htbp!]
    \centering
    \includegraphics[width=0.8\linewidth]{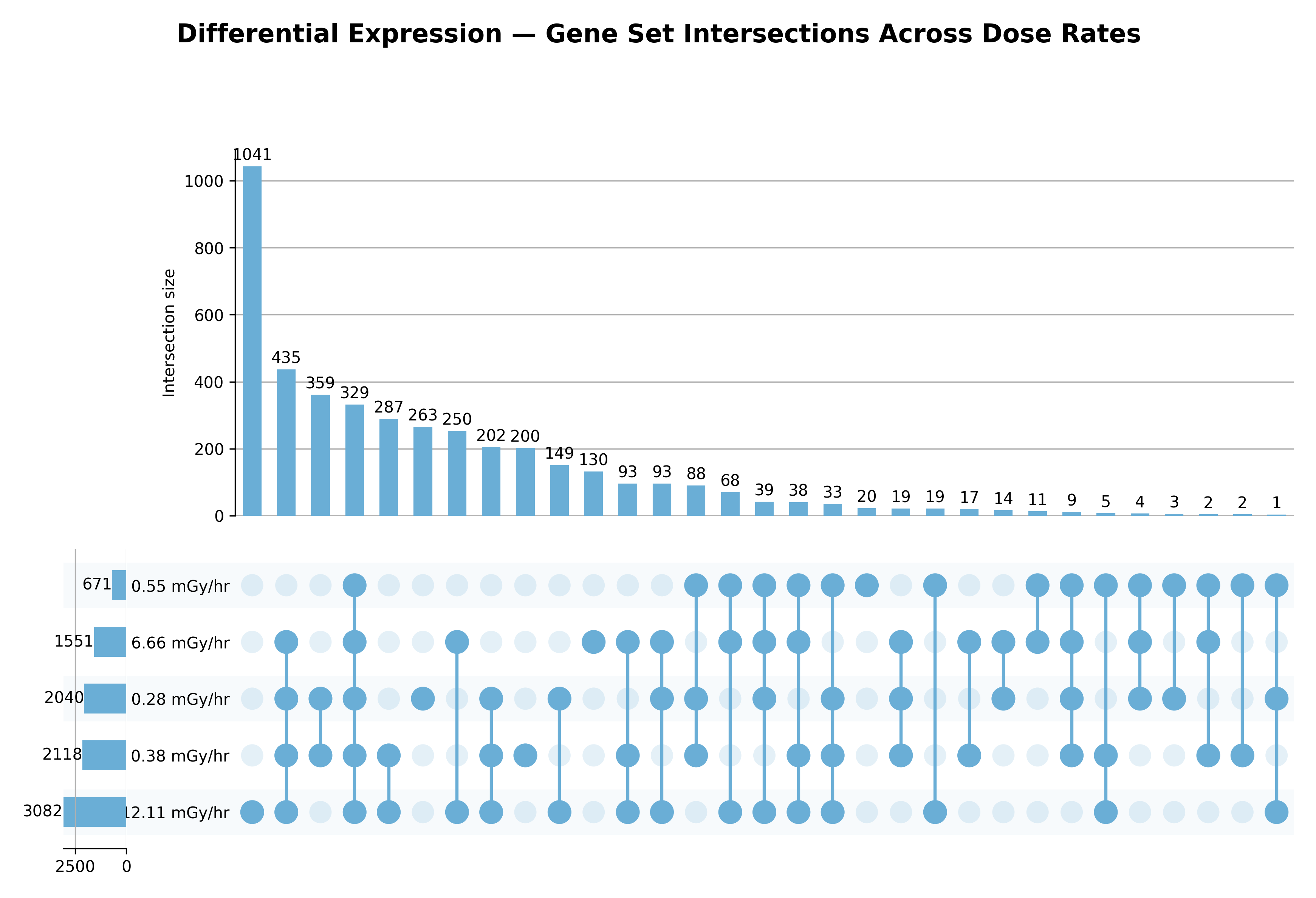}
    \caption{Intersections of differential expression graph sets across each dose rate. There is an \enquote{invariant} set of 329 genes which is the intersection across all dose rates.}
    \label{fig:upset_de}
\end{figure}

 There were 68 stable genes inferred by the \texttt{RandomForestRegression} models across folds. Intersections with causal gene sets and differential expression are shown in Fig. \ref{fig:venn_rf}. Fig. \ref{fig:upset_causal} and Fig. \ref{fig:upset_de}  show within causal graph and differential expression intersections respectively. There are a core set of \enquote{invariant} genes across dose rates for each method which may correspond to genes involved fundamental radiation response processes. The invariant gene set is larger for causal graphs than differential expression (438 genes compared to 329 genes).

\section{Random and Correlative Gene Sets }
\label{appendix:sanity_check}

\begin{figure}[H]
    \centering
    \includegraphics[width=\linewidth]{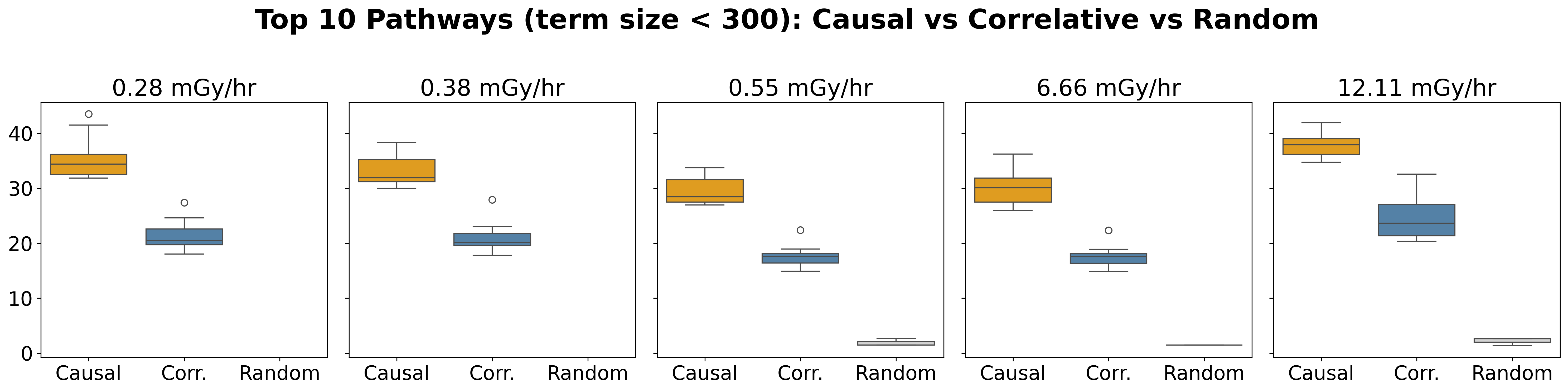}
    \caption{A sanity check for pathway enrichment checked enrichment of the top 10 pathways from the causal gene sets, correlative gene sets and random gene sets. }
    \label{fig:sanity_check}
\end{figure}

Pathway enrichment $-\text{log}_{10}p$-values were surprisingly large for causal gene sets (Fig. \ref{fig:pathway-enrichment-radiation}). To check that these values were meaningful, we ran a simple test with \texttt{gProfiler} with random genes taken from the background gene set (all genes measured in the experiment). In Fig. \ref{fig:sanity_check} we report the distribution of the top 10 enriched pathways (with 5 repeated runs). We verify random gene sets do not enrich any pathways. We also check our correlative linear model and see that these gene sets do enrich pathways, but not at the level of the causal gene sets. We already demonstrate in Fig. \ref{fig:pathway-enrichment-radiation} that the causal gene sets perform better compared differential expression and \texttt{RandomForest} gene sets. 

\section{Top 10 Pathways for Differential Expression Invariant Gene sets}
\label{appendix:top10_de}
We report the top 10 pathways for the differential expression invariant gene set (329 genes) in Table \ref{tab:invariant_top10_de_invariant}.
\begin{table}
\caption{Top 10 Enriched Pathways — Differential Expression Invariant Gene Set}
\label{tab:invariant_top10_de_invariant}
\centering
\begin{tabular}{p{6cm} r r r}
\toprule
Pathway & $-\log_{10}(p)$ & Term Size & Intersection \\
\midrule
\midrule
cellular response to interleukin-1 \newline {\small \texttt{GO:0071347}} & 9.73 & 84 & 13 \\
\hline
response to interleukin-1 \newline {\small \texttt{GO:0070555}} & 9.40 & 110 & 14 \\
\hline
cellular response to tumor necrosis factor \newline {\small \texttt{GO:0071356}} & 7.68 & 208 & 16 \\
\hline
glycosaminoglycan binding \newline {\small \texttt{GO:0005539}} & 7.15 & 249 & 17 \\
\hline
response to tumor necrosis factor \newline {\small \texttt{GO:0034612}} & 7.09 & 229 & 16 \\
\hline
glomerulus development \newline {\small \texttt{GO:0032835}} & 6.97 & 71 & 10 \\
\hline
active monoatomic ion transmembrane transporter activity \newline {\small \texttt{GO:0022853}} & 6.71 & 117 & 12 \\
\hline
Diabetic cardiomyopathy \newline {\small \texttt{KEGG:05415}} & 6.45 & 202 & 15 \\
\hline
regulation of angiogenesis \newline {\small \texttt{GO:0045765}} & 6.42 & 293 & 17 \\
\hline
nephron development \newline {\small \texttt{GO:0072006}} & 6.40 & 160 & 13 \\
\bottomrule
\end{tabular}
\end{table}

\section{Cluster Functional Analysis}
\label{appendix:cluster_analysis}

\begin{figure}[H]
    \centering
    \includegraphics[width=0.5\linewidth]{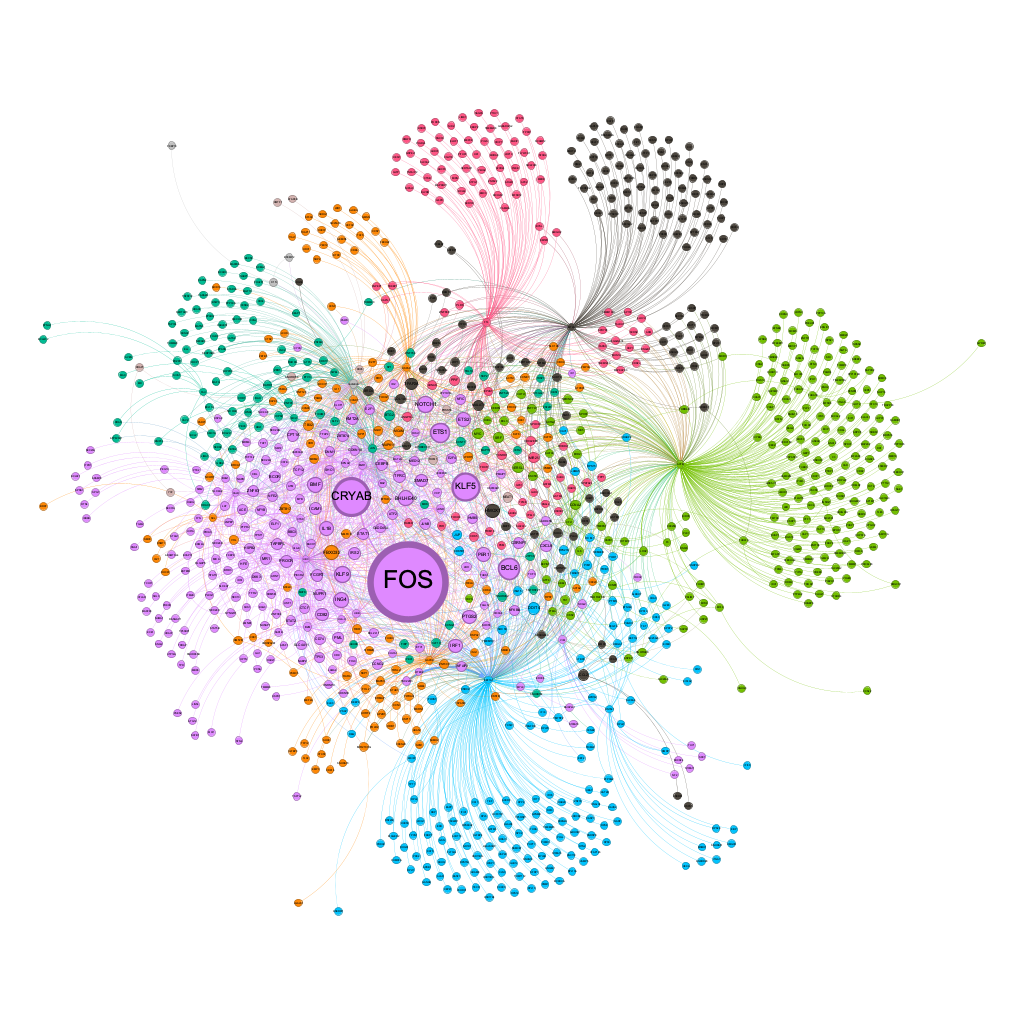}
    \caption{Example of the causal graph at dose rate 6.66 mGy/hr colored by cluster ID.}
    \label{fig:cluster_I}
\end{figure}

We performed a cluster analysis of the causal graphs at each dose rate; an example clustering is shown in Fig. \ref{fig:cluster_I}. To understand if each cluster is biologically meaningful and encodes distinct processes, we again perform pathway enrichment on each gene subset. We also compute overlaps  to CORUM protein complexes which are manually annotated, experimentally validated, high quality annotations. In Table \ref{tab:cluster_enrichment_I} we see significant enrichment of pathways and CORUM complexes in each cluster for dose rate 6.66 mGy/hr. Only C2 (Eukaryotic translation) and C3 (ATP- DNA binding), however, show clear consensus between the top pathway and the CORUM complex. 

\begin{table}
\centering
\caption{Cluster enrichment for dose rate I (6.66 mGy/min).
  CORUM complex overlap identified by Fisher's exact test;
  top pathway from gProfiler enrichment analysis.}
\label{tab:cluster_enrichment_I}
\scalebox{0.9}{
\begin{tabular}{lp{4.5cm}rp{4.5cm}r}
\toprule
Cluster & CORUM Complex & $p$-value & Top Pathway & $p$-value \\
\midrule
C0 & DNA binding; chromosome; DNA topological change & $4.81 \times 10^{-5}$ & Antigen Presentation: Folding, assembly and peptide loading of class I MHC & $8.10 \times 10^{-21}$ \\
C1 & Cell cycle checkpoint signaling; DNA binding; nucleus & $1.26 \times 10^{-4}$ & Pathogenic \textit{Escherichia coli} infection & $5.09 \times 10^{-7}$ \\
C2 & Cytoplasm; translation & $2.94 \times 10^{-24}$ & Eukaryotic Translation Elongation & $6.66 \times 10^{-22}$ \\
C3 & ATP binding; chromosome; DNA replication & $7.66 \times 10^{-10}$ & ATP-dependent activity, acting on DNA & $8.70 \times 10^{-9}$ \\
C4 & mRNA splicing, via spliceosome; spliceosomal complex & $1.93 \times 10^{-5}$ & Mitotic G1 phase and G1/S transition & $1.79 \times 10^{-16}$ \\
C5 & Protein targeting; intracellular protein transport; endocytosis & $1.82 \times 10^{-2}$ & NRF2 pathway & $3.91 \times 10^{-6}$ \\
C6 & Myeloid cell differentiation; regulation of apoptotic process & $9.58 \times 10^{-9}$ & Integrated breast cancer pathway & $9.83 \times 10^{-10}$ \\
\bottomrule
\end{tabular}
}
\end{table}

\section{Causal Discovery Bootstrap Results}
\label{appendix:bootstrap}
\begin{figure}[htpb!]
    \centering
    \begin{minipage}[t]{0.51\textwidth}
        \centering
        \includegraphics[width=\linewidth]{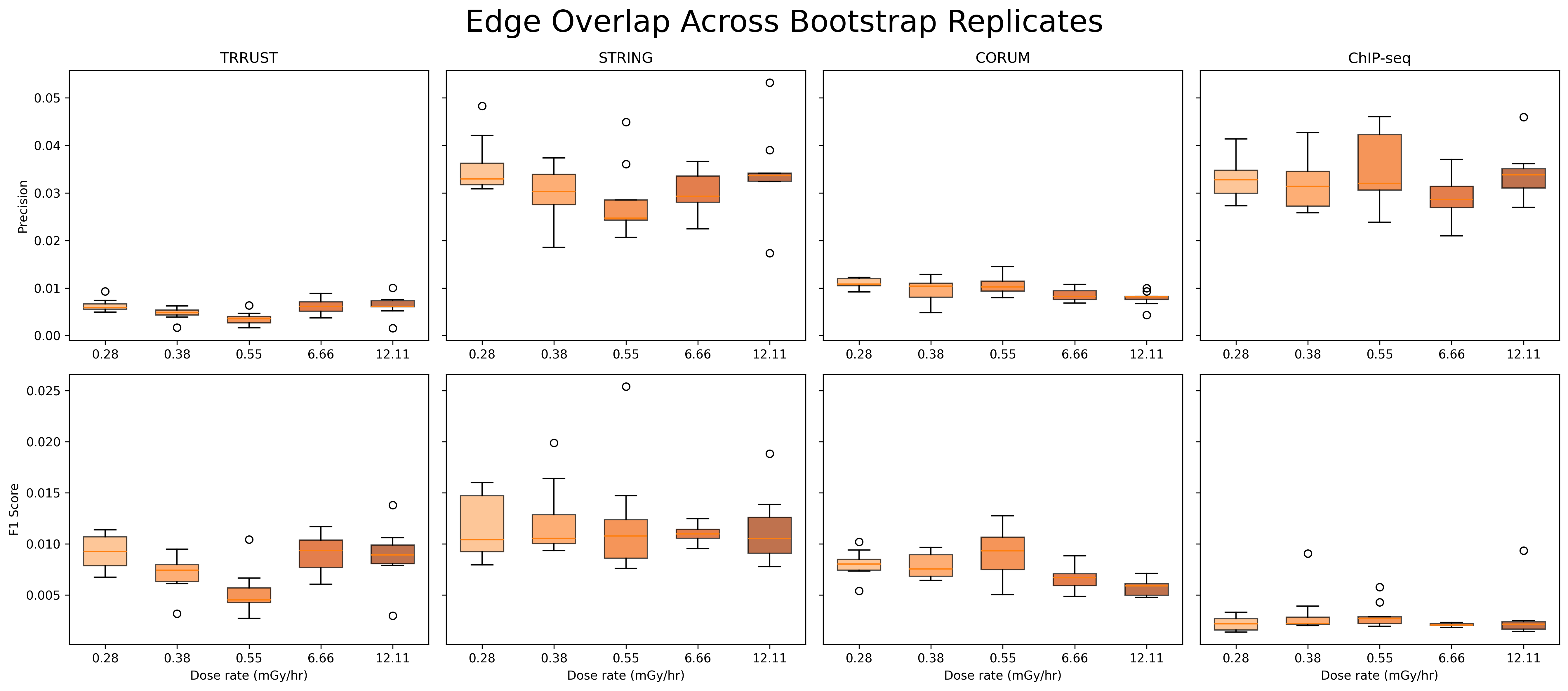}
        \label{fig:boot_edge}
    \end{minipage}
    \hfill
    \begin{minipage}[t]{0.45\textwidth}
        \centering
        \includegraphics[width=\linewidth]{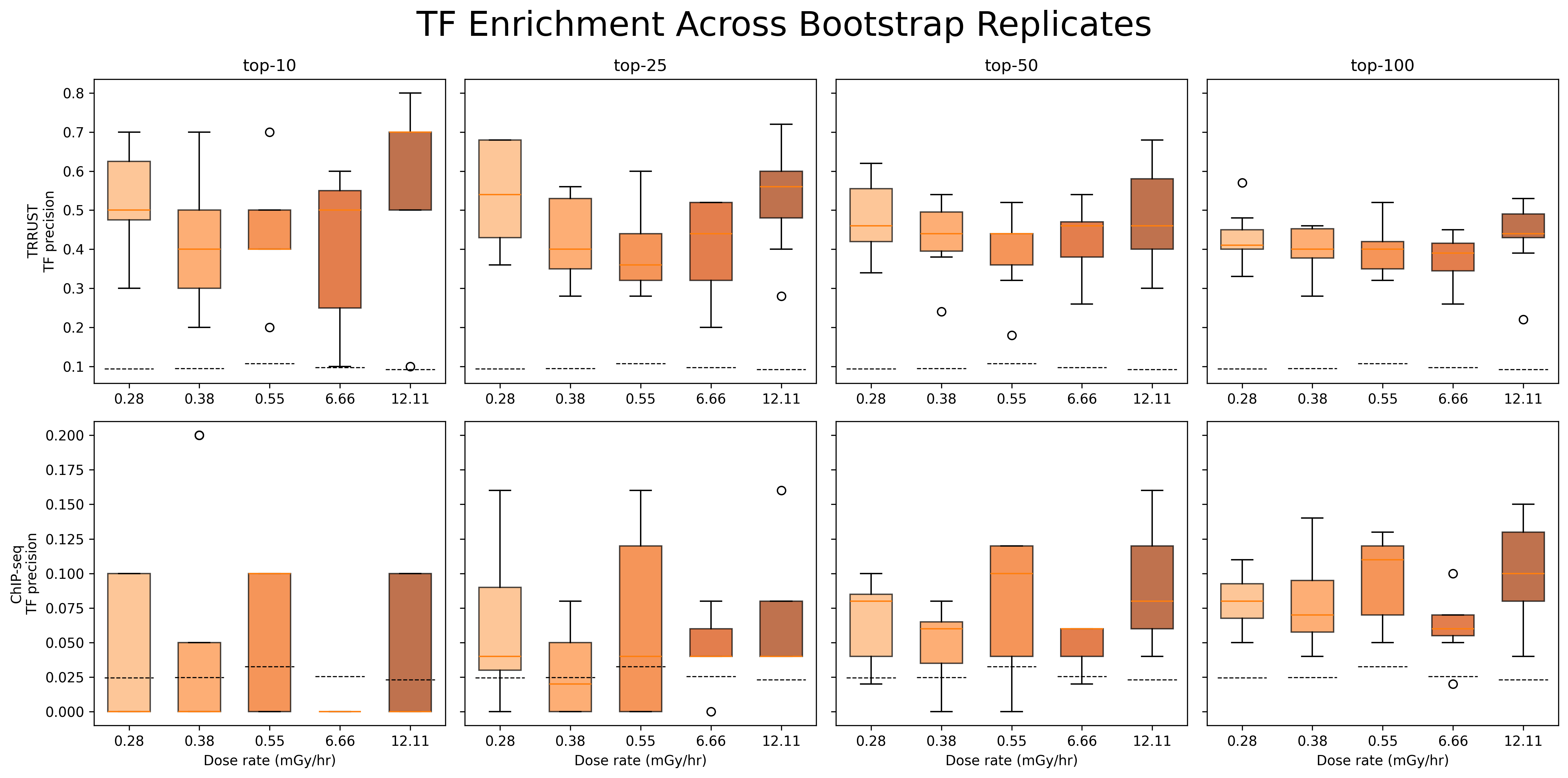}
        \label{fig:boot_tf}
    \end{minipage}

    \caption{(\textbf{Left}) The distribution of edge overlap with knowledge bases for all graphs across 10 bootstrap runs. (\textbf{Right}) The percentage of top-k hub nodes that correspond to known transcription factors for all graphs in the 10 bootstrap runs.}
    \label{fig:boot_metrics}
\end{figure}

\begin{figure}[H]
    \centering
    \includegraphics[width=\linewidth]{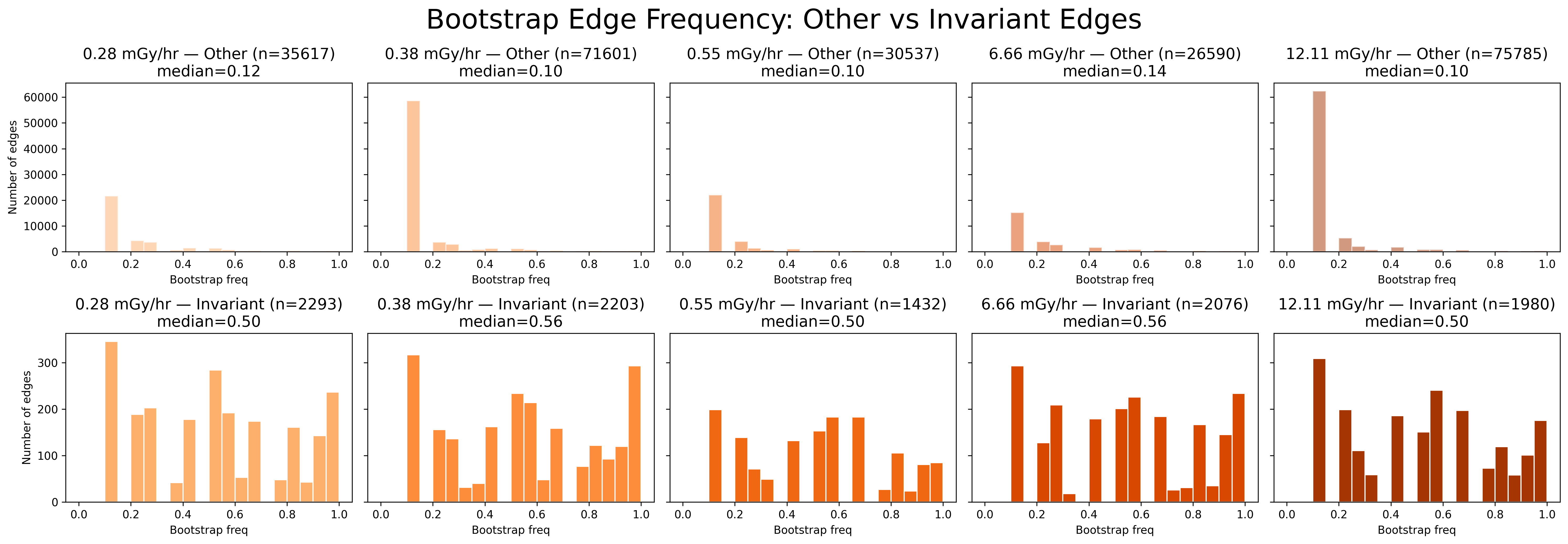}
    \caption{(\textbf{Top}) The distribution of edge frequencies at each dose rate across bootstraps. A frequency of 1.0 means that the edge appeared in all 10 of the DAGs. (\textbf{Bottom}) The distribution of edge frequencies at each dose rate subgraph induced by the causal invariant gene set (438 genes). }
    \label{fig:boot_invariant}
\end{figure}

Here, we show the \texttt{DAG-GNN} metrics across bootstrap runs. The results in the main paper report metrics for consensus graphs formed by filtering for edges that co-occur in 50\% or more of the bootstrap runs. We ran \texttt{DAG-GNN} with 10 bootstraps and metrics are shown in Fig. \ref{fig:boot_metrics}. Compared to Table \ref{tab:edge_overlap} and Fig. \ref{fig:hub_tf_enrichment}, we see a high variance distribution and overall poorer results which supports our choice of choosing consensus edges for downstream analysis. In Fig. \ref{fig:boot_invariant}, we visualize the distribution of edge frequencies for subgraphs induced by the invariant causal gene set. We see that this subgraph has a higher density of \enquote{perfect} edges that appear in all bootstraps. This shows that the invariant gene sets and subgraph edges have a higher confidence, and this is supported by enrichment of only radiation pathways. These subgraphs and high frequency edges can be used as a starting point for hypothesized novel pathways, however, this requires experiments to prove. 

\end{document}